\documentclass[aps,prl,twocolumn,nofootinbib,superscriptaddress,preprintnumbers,floatfix]{revtex4-2}

\usepackage{amsmath,amssymb}
\usepackage{bm}
\usepackage{graphicx}
\usepackage[hypertexnames=false]{hyperref}

\begin{document}

\preprint{\vbox{\hbox{MIT-CTP/5579}}}
\title{Two-pole nature of the \texorpdfstring{$\Lambda(1405)$}{Lambda(1405)} from lattice QCD} 

\author{John Bulava}
\affiliation{Deutsches Elektronen-Synchrotron (DESY), Platanenallee 6, 15738 Zeuthen, Germany}

\author{B\'{a}rbara Cid-Mora}
\affiliation{GSI Helmholtz Centre for Heavy Ion Research, 64291 Darmstadt, Germany}

\author{Andrew D. Hanlon}
\affiliation{Physics Department, Brookhaven National Laboratory, Upton, New York 11973, USA}

\author{Ben H\"{o}rz}
\affiliation{Intel Deutschland GmbH, Dornacher Str. 1, 85622 Feldkirchen, Germany}

\author{Daniel Mohler}
\affiliation{Institut f\"ur Kernphysik, Technische Universit\"at Darmstadt,
Schlossgartenstrasse 2, 64289 Darmstadt, Germany}
\affiliation{GSI Helmholtz Centre for Heavy Ion Research, 64291 Darmstadt, Germany}

\author{Colin Morningstar}
\affiliation{Department of Physics, Carnegie Mellon University, Pittsburgh, Pennsylvania 15213, USA}

\author{Joseph Moscoso}
\affiliation{Department of Physics and Astronomy, University of North Carolina, Chapel Hill, NC 27516-3255, USA}

\author{Amy Nicholson}
\affiliation{Department of Physics and Astronomy, University of North Carolina, Chapel Hill, NC 27516-3255, USA}

\author{Fernando Romero-L\'{o}pez}
\affiliation{Center for Theoretical Physics, Massachusetts Institute of Technology, Cambridge, MA 02139, USA}

\author{Sarah Skinner}
\affiliation{Department of Physics, Carnegie Mellon University, Pittsburgh, Pennsylvania 15213, USA}

\author{Andr\'{e} Walker-Loud}
\affiliation{Nuclear Science Division, Lawrence Berkeley National Laboratory, Berkeley, CA 94720, USA}

\collaboration{for the Baryon Scattering (BaSc) Collaboration}

\date{\today}

\begin{abstract}
This letter presents the first lattice QCD computation of the coupled channel $\pi\Sigma$--$\bar{K}N$ scattering 
amplitudes at energies near $1405\,{\rm MeV}$. These amplitudes contain the resonance $\Lambda(1405)$ with 
strangeness $S=-1$ and isospin, spin, and parity quantum numbers $I(J^P)=0(1/2^-)$. However, whether there is a 
single resonance or two nearby resonance poles in this region is controversial theoretically and experimentally.
Using single-baryon and meson-baryon operators to extract the finite-volume stationary-state energies to obtain 
the scattering amplitudes at slightly unphysical quark masses corresponding to $m_\pi\approx200$~MeV and 
$m_K\approx487$~MeV, this study finds the amplitudes exhibit a virtual bound state below the $\pi\Sigma$ 
threshold in addition to the established resonance pole just below the $\bar{K}N$ threshold. Several 
parametrizations of the two-channel $K$-matrix are employed to fit the lattice QCD results, all of which 
support the two-pole picture suggested by $SU(3)$ chiral symmetry and unitarity.  
\end{abstract}


\nopagebreak
\maketitle

{\it Introduction.}---The strong nuclear force is described by quantum chromodynamics (QCD) which governs
the dynamics and interactions of quarks and gluons.  Due to an important property of QCD known as asymptotic
freedom, the use of perturbation theory is useful for QCD scattering calculations at very high energies.
The binding of quarks and gluons to form hadrons, such as protons and neutrons, is a low-energy phenomenon of
QCD, requiring a nonperturbative calculational technique. Such techniques are difficult to apply, so that
understanding the hadron spectrum of QCD remains an important outstanding issue for the Standard Model of
particle physics. In particular, resonances such as the $\Lambda(1405)$ defy the naive quark-model picture
of baryons and mesons. In this letter, a Markov-chain Monte Carlo method using QCD formulated on a
space-time lattice is applied to shed light on the puzzling hadron resonance structure in the region of the
$\Lambda(1405)$.

The history of the $\Lambda(1405)$ began in Refs.~\cite{Dalitz:1959dn,Dalitz:1960du} which
suggested that the low-energy $K^-p$ amplitude measured in bubble chamber experiments implies a resonance
in the $\pi^{-} \Sigma^{+}$ spectrum just below the $K^{-}p$ threshold. The intervening decades have
witnessed considerable experimental progress in this system~\cite{Hyodo:2011ur, Hyodo:2020czb}, but a
consensus about whether there is a single resonance or two nearby resonance poles in this energy region
has not yet been reached. This is evidenced by the most recent Particle Data Group
review~\cite{ParticleDataGroup:2022pth} which lists an additional $\Lambda(1380)$
resonance pole with lower confidence. An improved determination of the $K^-p$ scattering
length~\cite{Meissner:2004jr} was enabled by measurements of the energy shift and width of kaonic hydrogen
by the SIDDHARTHA collaboration at DA$\Phi$NE~\cite{SIDDHARTA:2011dsy}.  The angular analysis of the process
$\gamma + p \rightarrow K^{+} + \Sigma + \pi$ by the CLAS collaboration at JLab determined the line
shapes~\cite{CLAS:2013rjt} and the spin parity quantum numbers~\cite{PhysRevLett.112.082004} $J^{P} =1/2^{-}$.
The CLAS data has been analyzed in Refs.~\cite{Mai:2014xna,Roca:2013cca}, which suggest the existence of two
isoscalar poles. Recent data from the BGOOD collaboration~\cite{BGOOD:2021sog} and a preliminary study
by the GlueX collaboration~\cite{Wickramaarachchi:2022mhi} also support the two-pole scenario. Similarly,
inter-hadron potentials determined by the ALICE collaboration using the `femtoscopy' approach favor the
two-pole picture~\cite{ALICE:2022yyh}. However, recent data from J-PARC is successfully described by a
single pole~\cite{J-PARCE31:2022plu} and a combined analysis in Ref.~\cite{Anisovich:2020lec} concludes
that a single resonance sufficiently describes the experimental data, without ruling out the two-pole
description.

On the theoretical side, the relatively low mass and quantum numbers of the $\Lambda(1405)$ are difficult to
accommodate in constituent quark models~\cite{Isgur:1978xj}. However, some insight is gained from
$SU(3)$ chiral effective theory~\cite{Kaiser:1995eg,Oset:1997it,MartinezTorres:2012yi}. The leading-order
interaction between the octet of Goldstone bosons and (ground-state) octet
baryons~\cite{Weinberg:1966kf,Tomozawa:1966jm} predicts an attractive interaction in both the flavor-$SU(3)$
singlet and octet combinations. After employing a unitarization procedure, this attraction leads to two poles
in the scattering matrix analytically continued to complex center-of-mass energies~\cite{Oller:2000fj}.
Despite the agreement of nearly all chiral approaches (which are reviewed in
Refs.~\cite{Mai:2018rjx,Mai:2020ltx,Meissner:2020khl}) on the two-pole scenario, the
position of the lower pole remains somewhat poorly constrained~\cite{ParticleDataGroup:2022pth}.
Recent theoretical works about the $\Lambda(1405)$ can be found in
Refs.~\cite{Ezoe:2020piq,Myint:2018ypc,Miyahara:2018onh,Miyahara:2018lud,Azizi:2023tmw,Xie:2023cej,Lu:2022hwm,Hyodo:2022xhp}.

Lattice QCD is a first-principles method that can be used to unambiguously determine the nature of the 
$\Lambda(1405)$ and provide two unique insights. First, the elastic $\pi\Sigma$ scattering amplitude can be computed
directly below the $\bar{K}N$ threshold. This process is difficult to access experimentally and lattice results may 
help identify and constrain a second lower pole. Second, the motion of the poles in the complex plane upon varying 
the $u$, $d$, and $s$ quark masses away from their physical values provides additional input to future chiral 
effective theory analyses~\cite{Jido:2003cb,Molina:2015uqp}. 

The computation of real-time two-to-two scattering amplitudes below three-hadron thresholds from imaginary-time 
lattice QCD calculations is well-developed and relies on the finite-volume spectrum of interacting two-hadron 
states~\cite{Luscher:1990ux,Rummukainen:1995vs, Kim:2005gf, He:2005ey, Bernard:2010fp, Gockeler:2012yj,
Briceno:2012yi, Briceno:2014oea}. Previous lattice QCD computations of the $\Lambda(1405)$ have not computed 
scattering amplitudes and instead aimed only to isolate the lowest finite-volume energy eigenstate using 
single-baryon three-quark interpolating fields~\cite{Gubler:2016viv,Menadue:2011pd,Engel:2012qp,Engel:2013ig,
Nemoto:2003ft,Burch:2006cc,Takahashi:2009bu,Meinel:2021grq,Hall:2014uca}. Using only such operators is known 
to be insufficient to extract scattering information, such as scattering amplitudes and pole locations.
The $\bar{K}N$ scattering length for $I=0$ has been computed long ago using the quenched
approximation~\cite{Fukugita:1994ve}, but mixing with the kinematically-open $\pi\Sigma$ channel was neglected.
The $\pi\Sigma$ and $\bar{K}N$ scattering lengths in other (non-singlet) flavor and isospin combinations not
directly relevant for the $\Lambda(1405)$ have been computed 
in Refs.~\cite{Detmold:2015qwf,Torok:2009dg,Meng:2003gm}. 

This work computes the isospin $I=0$ and strangeness $S=-1$ coupled-channel $\pi\Sigma-\bar{K}N$ scattering amplitudes 
below the $\pi\pi\Lambda$ threshold from lattice QCD for the first time. A single ensemble of gauge configurations 
with dynamical $u$, $d$, and $s$ quarks is employed with pion and kaon masses of $m_{\pi} \approx 200\,{\rm MeV}$ and 
$m_{\rm K}\approx487$~MeV, respectively, which deviate slightly from their physical values 
$m_\pi^{\rm phys}\approx 140$~MeV and $m_{\rm K}^{\rm phys}\approx 495$~MeV.  The $u$ and $d$ quark masses are set to
be equal and electroweak interactions are neglected, so isospin is a good quantum number. The main result of this 
work is a set of parametrizations of the amplitudes which are constrained by fits to the finite-volume energy spectrum. 
These parametrizations can accommodate zero, one, or two poles,  but when fit to the lattice results and analytically 
continued to the complex-energy plane, they all confirm the presence of two poles, the positions of which vary little 
and are consistent with predictions from chiral effective theory.  Our use of $m_\pi > m_\pi^{\rm phys}$ moves the 
lower pole just below the $\pi\Sigma$-threshold leading to its unambiguous identification as a virtual bound state. 
The higher pole near the $\bar{K}N$-threshold is also clearly present.

\begin{figure}[t]
\includegraphics[width=\linewidth]{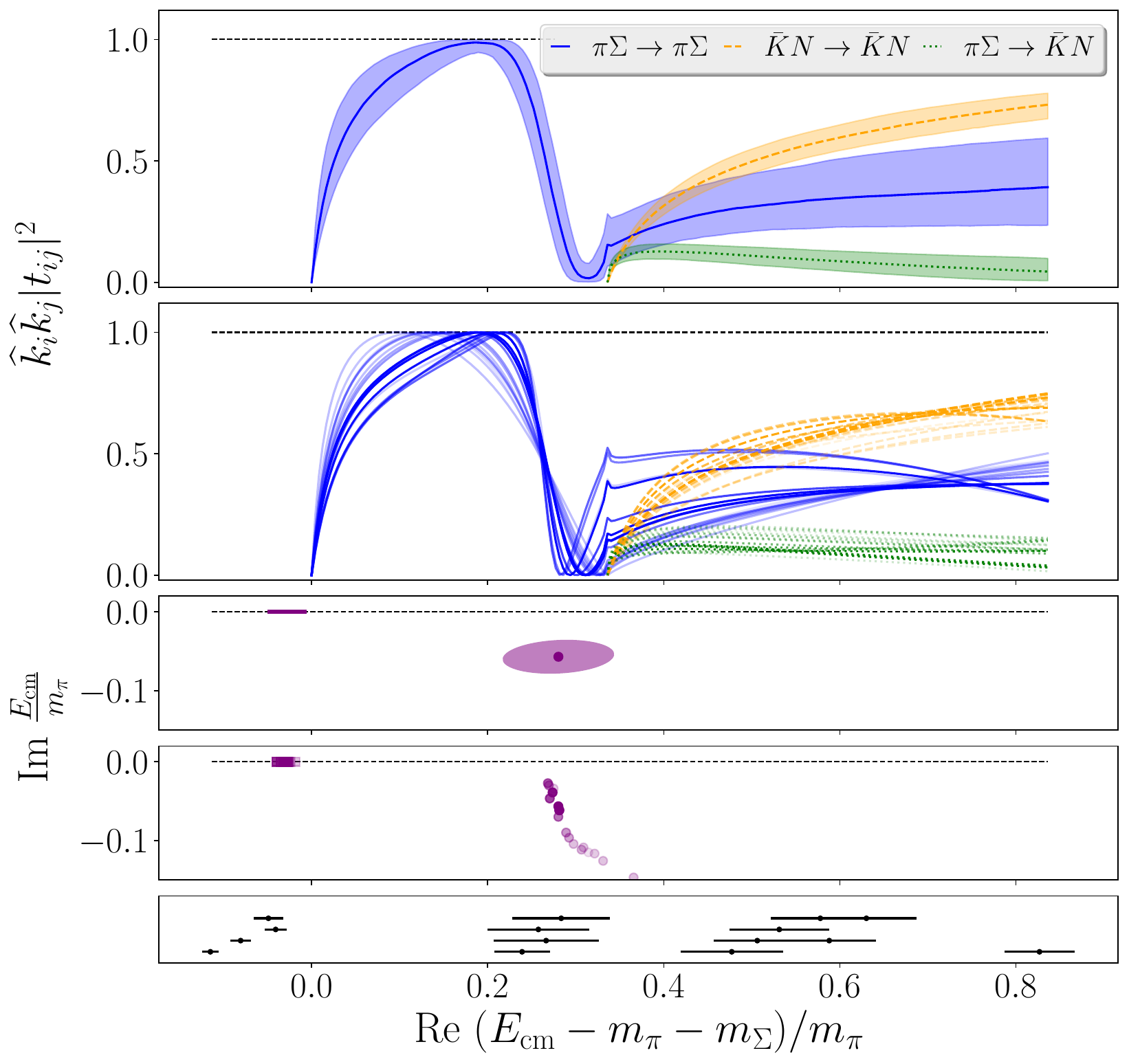}
\caption{The $I=0$ and $S=-1$ coupled-channel $\pi\Sigma-\bar{K}N$ amplitude computed on a single lattice
QCD gauge-field ensemble with $m_{\pi} \approx 200\,{\rm MeV}$ as a function of the energy difference to the
$\pi\Sigma$ threshold in the center-of-mass frame.  The upper panel shows the transition matrix elements, defined
in Eq.~(\ref{eq:amplitude}), using the $K$-matrix parametrization with the lowest AIC constrained by the
finite-volume spectrum in the bottom panel. The second  panel  shows the model variation for the same quantities
using several parametrizations. The third and fourth panels show the position of poles in the complex
center-of-mass energy ($E_{\rm cm}$) plane on the sheets closest to the physical one: using the parametrization
with lowest AIC (third panel), and for several parametrizations (fourth panel). In the second and fourth panel,
the transparency of each line and corresponding pair of pole positions is proportional to
$\exp{[- \left(\text{AIC} - \text{AIC}_\text{min}\right)/2]}$, where $\text{AIC}_\text{min}$ is the lowest AIC
corresponding to the  fit in Eq.~(\ref{eq:bestfit}), which is also shown in the top panel.  The subscripts $i,j$
index the two open scattering channels. In the lowest panel, the lattice QCD energy levels that serve as input
to the amplitude analyses are displayed. For clarity, these energy levels are displaced
vertically by the total spatial momentum $\boldsymbol d^2$ defined below Eq.~(\ref{e:det}).
\label{fig:main}}
\end{figure}

This letter provides a summary of the computation while technical details are left to a companion 
paper~\cite{BaryonScatteringBaSc:2023ori}. 
The main result is Fig.~\ref{fig:main}, which shows fits to the finite-volume spectrum using all parametrizations 
of the coupled-channel amplitude and the associated pole positions. Statistical errors are shown for the 
parametrization with the lowest Akaike information criterion (AIC) value.  

{\it Determination of finite-volume energies.}--- 
The ensemble of gauge configurations and algorithm for evaluating correlation functions are briefly reviewed 
here and discussed more deeply in the companion paper. The $N_{\rm f} = 2+1$ QCD gauge configurations comprise 
the `D200' ensemble generated by the Coordinated Lattice Simulations (CLS) consortium~\cite{Bruno:2014jqa} which 
is detailed in Table~\ref{tab:comp_deets}. The lattice spacing is determined in Ref.~\cite{Bruno:2016plf} and 
updated in Ref.~\cite{Strassberger:2021tsu}.  All correlation matrices are computed using the 
stochastic-LapH~\cite{Morningstar:2011ka} implementation of Ref.~\cite{Bulava:2022vpq}. The flexibility afforded 
by the source-sink factorization and subsequent computation of correlators via optimized tensor 
contractions~\cite{Horz:2019rrn} is particularly advantageous for large Hermitian correlation matrices containing 
single-baryon, $\pi\Sigma$, and $\bar{K}N$ interpolating operators.

\begin{table}[t]
\caption{
Parameters of the D200 ensemble~\cite{Bruno:2014jqa}. The lattice dimensions in space and time ($L$ and $T$), as 
well as the mass of the pion ($m_{\pi}$) and kaon ($m_{\rm K}$) are given in units of the lattice spacing $a$.
In pion mass units, the size of the box is $m_\pi L = 4.181(16)$.
\label{tab:comp_deets}}
\begin{ruledtabular}
\begin{tabular}{c@{\hskip 12pt}c@{\hskip 12pt}c@{\hskip 12pt}c@{\hskip 12pt}}
 $a [\textup{fm}]$ & $(L/a)^{3} \times T/a$ & $am_{\pi}$ & $am_{\rm K}$  \\ \hline
  0.0633(4)(6) & $64^{3} \times 128$ & 0.06533(25) &  0.15602(16)
\end{tabular}
\end{ruledtabular}
\end{table}

\begin{figure}[t]
\centering
\includegraphics[width=\linewidth]{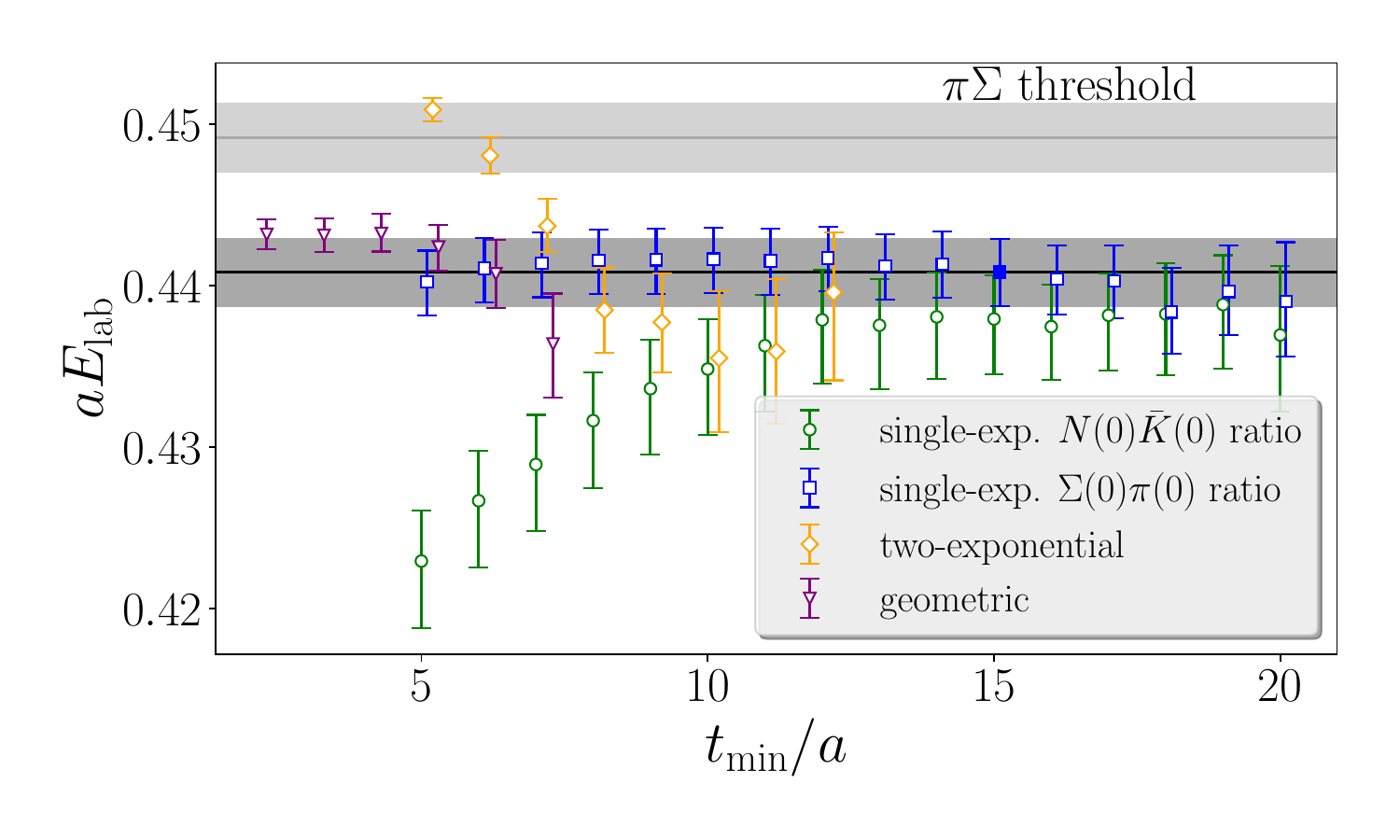}
\caption{Example determination of a finite-volume stationary-state energy, illustrating 
the sensitivity of the fitted energy to the lower end of the fit range ($t_{\rm min}$) for the lowest level of 
the $G_{\mathrm{1u}}(0)$ irrep. Each set of points corresponds to a different fit form. The two-exponential 
and geometric~\cite{Bulava:2022vpq} fits are performed to the diagonalized correlation function only. 
The single-exponential ratio fits are performed to the same correlator divided by either the product 
$\bar{K}(0)N(0)$ or $\pi(0)\Sigma(0)$ of correlators, and the lab frame energy $aE_{\rm lab}$ is reconstructed 
from the interaction shifts. The dark horizontal band and filled symbol denote the chosen fit.
\label{fig:g1u_example} }
\end{figure}

\begin{figure}
\centering
\includegraphics[width=\linewidth]{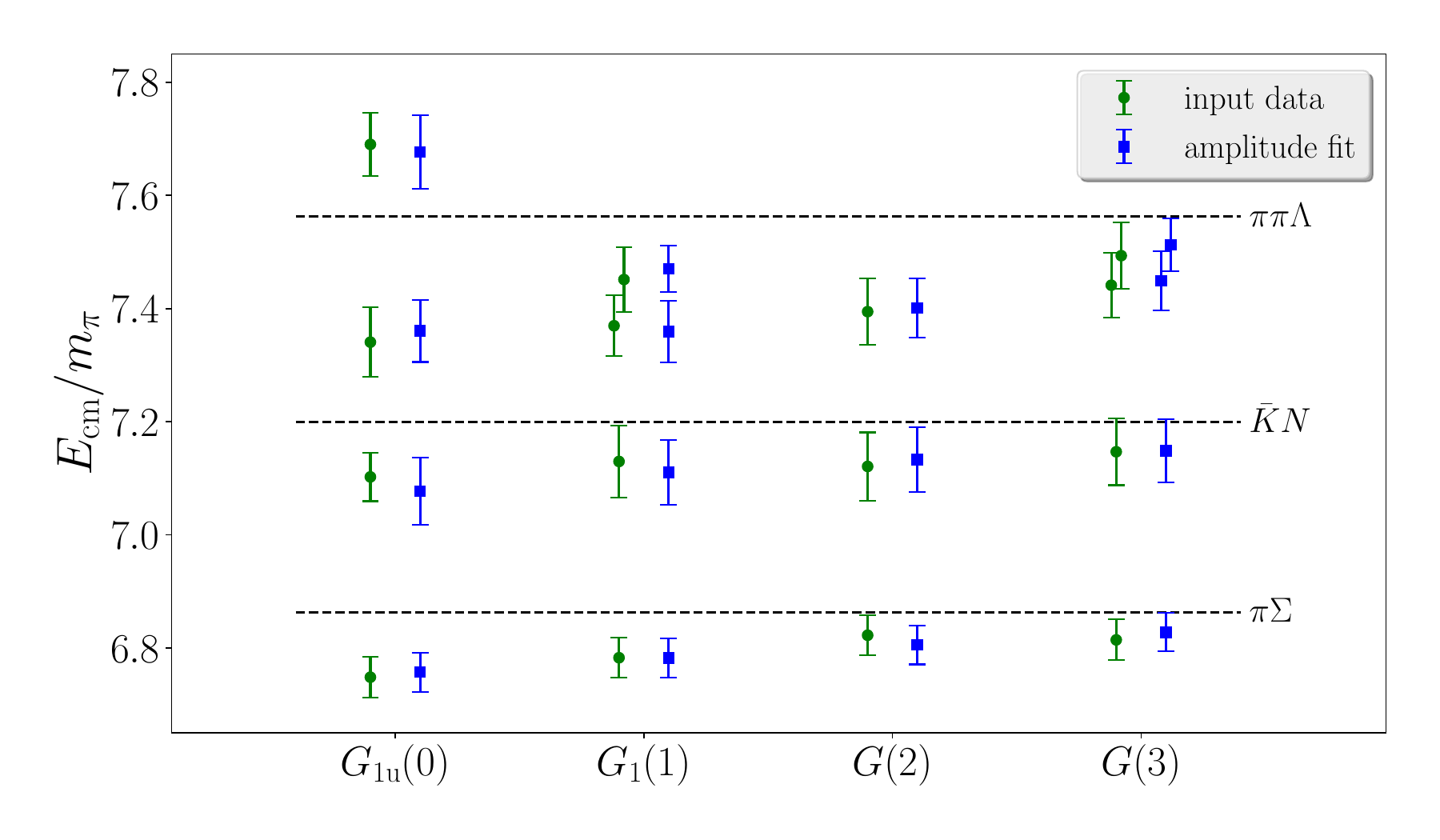}
\caption{Finite-volume spectrum in the center-of-mass frame used as input data to constrain 
parametrizations of the coupled-channel $\pi\Sigma-\bar{K}N$ scattering amplitude.  Each column 
corresponds to a particular irrep $\Lambda(\boldsymbol{d}^2)$ of the little group of total momentum 
$\boldsymbol{P}^2=(2\pi/L)^2\boldsymbol{d}^2$. Only irreps where the $\ell=0$ partial wave contributes 
are included.  Dashed lines indicate various thresholds, as labeled. Model energies from the 
resultant scattering-amplitude fit are given by blue squares.
\label{fig:energy_spectrum}}
\end{figure}

The determination of finite-volume stationary-state energies is also discussed in detail in the companion paper 
and summarized here.  The interaction shift $\Delta E_{\rm lab}$ of a lab-frame energy from a nearby 
non-interacting energy  is extracted from a single-state fit to the ratio of a diagonalized correlation function 
over the product of correlators for the individual constituents of the nearby non-interacting energy.  The 
diagonalization of the correlation matrices is done by solving a generalized eigenvalue problem (GEVP) as 
described in Ref.~\cite{Bulava:2022vpq}. We have verified the insensitivity of our extracted energies to 
reasonable variations of the GEVP parameters, the use of different nearby non-interacting levels, and different 
fit forms. An example energy determination is shown in Fig.~\ref{fig:g1u_example} for the ground state of the 
$G_{\rm 1u}(0)$ irreducible representation (irrep), which predominantly contains the parity-odd, $s$-wave 
scattering system. All levels used to constrain the amplitude are shown in Fig.~\ref{fig:energy_spectrum}.

{\it Scattering amplitude determination.}--- 
In lattice QCD, scattering amplitudes below three-hadron thresholds are inferred from finite-volume 
spectra~\cite{Luscher:1990ux,Rummukainen:1995vs, Kim:2005gf, He:2005ey, Bernard:2010fp, Gockeler:2012yj,
Briceno:2012yi, Briceno:2014oea} using the relationship~\cite{Morningstar:2017spu}
\begin{align}\label{e:det}
    {\rm det}[ \widetilde{K}^{-1}(E_{\rm cm}) - B^{\boldsymbol{P}}(E_{\rm cm})]  = 0 \,.
\end{align}
The matrix $\widetilde{K}$ is related to
the usual scattering $K$-matrix (normalized such that the single-channel equivalent of $ \widetilde{K}^{-1}$
is the $s$-wave $k \cot \delta_0$), and the `box matrix' $B^{\boldsymbol{P}}$ for a particular total momentum
$\boldsymbol{P} = (2\pi/L) \boldsymbol{d}$ (with $\boldsymbol{d} \in \mathbb{Z}^3$) encodes the reduction in
symmetry due to the finite toroidal spatial volume. $E_{\rm cm}$ denotes center-of-mass energy.
Eq.~(\ref{e:det}) ignores terms which are suppressed exponentially with the spatial extent $L$. For
scattering between baryons and pseudoscalar mesons, the $K$-matrix does not mix different $J^P$, but does
couple the $\pi\Sigma$ and $\bar{K}N$ channels. By contrast, the box matrix is diagonal in the two scattering
channels, but mixes partial waves. $B^{\boldsymbol{P}}$ is, however, block diagonal in the basis given by
irreps of the finite-volume little group of momentum $\boldsymbol{P}$. Demanding a vanishing  determinant in
one of these infinite-dimensional blocks  provides a relationship between the $K$-matrix and the
finite-volume spectrum in a particular irrep. In practice, partial waves with orbital angular momentum
$\ell > \ell_{\rm max}$ are neglected; here $\ell_{\rm max} = 0$  is chosen in both the $\pi\Sigma$ and
$\bar{K}N$ channels.  The systematic error due to this is estimated by considering $\ell_{\rm max} = 1$,
and found to be insignificant for the near-threshold energy region relevant here. Specifically, the effect
of including additional waves with $\ell=1$ leads to shifts that are significantly smaller that the
statistical uncertainties in fit results for the $\ell=0$ $K$-matrix. Levels from all irreps in Table~1 of
Ref.~\cite{Bulava:2022vpq} to which the $J^P=1/2^-$ partial wave contributes are employed, as well as one
level each from the $G_{\rm 1g}(0)$, $F_1(3)$, and $F_2(3)$ irreps for the $\ell_{\rm max}=1$ check.
All elements of $B^{\boldsymbol{P}}$ required for this work are given in Ref.~\cite{Morningstar:2017spu}.

For $\ell_{\rm max} = 0$, the finite-volume spectrum shown in Fig.~\ref{fig:energy_spectrum} 
constrains the coupled-channel scattering amplitude via Eq.~(\ref{e:det}) at center-of-mass 
energies near the $\pi\Sigma$ and $\bar{K}N$ thresholds. The effective range expansion (ERE) 
is used to parametrize the $K$-matrix 
\begin{equation}
   \frac{E_{\rm cm}}{m_\pi} \widetilde{K}_{ij} = A_{ij} + B_{ij} \Delta_{\pi \Sigma},
   \label{eq:parametrization}
\end{equation}
where $A_{ij}$ and $B_{ij}$ are symmetric and real coefficients with $i$ and $j$ denoting either 
of the two scattering channels (channel 0 is $\pi\Sigma$ and channel 1  is $\bar KN$). Moreover,  
$\Delta_{\pi \Sigma} = (E_{\rm cm}^{2} - (m_{\pi} + m_{\Sigma})^{2})/(m_{\pi} + m_{\Sigma})^{2}$ 
labels the distance to the $\pi\Sigma$ threshold. The parameters, which are the elements of the 
$A$ and $B$ matrices, are determined from fits to the lattice QCD results using the spectrum 
method~\cite{Guo:2012hv}. Similar fits are performed with variations of the above parametrization: 
an ERE for $\widetilde{K}^{-1}$, removing the factor of $E_{\rm cm}$ in Eq.~(\ref{eq:parametrization}),
parametrizations inspired by the Weinberg-Tomozawa potential~\cite{Oset:1997it}, 
or using the Blatt-Biedenharn~\cite{Blatt:zz1952a} form. The effect of fixing some (or all) of the 
elements of $B$ to zero is also explored.

The correlated-$\chi^2$ of the above fits is defined by comparing the center-of-mass interaction 
shifts $\Delta E_{\rm cm}$ obtained from the model with those determined from the ratio fits with 
a particular choice of the non-interacting levels.  The fit with the lowest Akaike Information 
Criterion (AIC) value is a four-parameter fit to Eq.~(\ref{eq:parametrization}). The result is
\begin{align}
\begin{split}
    &A_{00} = 4.1(1.8), \quad \ \, A_{11}=-10.5(1.1), \\ &A_{01}=10.3(1.5), \quad B_{01}=-29(18), 
    \label{eq:bestfit}
\end{split}
\end{align}
with fixed $B_{00}=B_{11}=0$ and $\chi^{2}=10.52$ for 11 degrees of freedom. This fit is shown in 
Fig.~\ref{fig:main}. All statistical uncertainties and correlations are taken into account using 
the bootstrap method with $800$ samples.

{\it Analytic structure of the amplitude.}---The various parametrizations discussed above constrain 
the energy dependence of the amplitudes near the finite-volume energies, even if they do not 
accommodate left-hand (cross-channel) cuts. Knowledge over this limited range enables the analytic 
continuation of the scattering amplitude (denoted $\mathcal{T}$) to complex $E_{\rm cm}$ and the 
identification of poles close to the real axis on sheets adjacent to the physical one. 

The $K$-matrix, the $J^{P}=1/2^{-}$ scattering amplitude $\mathcal T$, and the normalized amplitude 
$t$ shown in Fig.~\ref{fig:main} are related by
\begin{equation}
t^{-1} = \frac{8\pi E_{\rm cm}}{m_\pi}\, \mathcal T^{-1} =    \widetilde{K}^{-1} -i \widehat k, \quad 
\label{eq:amplitude}
\end{equation}
where 
$m_\pi\widehat k = \text{diag }(k_{\pi \Sigma}, k_{\bar{K}N})$,
\begin{align*}
k_{\pi \Sigma}^{2} &= \frac{1}{4E^{2}_{\rm cm }}\lambda(E^{2}_{\rm cm}, m_{\pi}^{2}, m_{\Sigma}^{2})\, .
\end{align*}
Here, $\lambda(x,y,z)$ is the K\"all\'en function~\cite{KallenBook} and $k_{\bar{K}N}$ is defined 
similarly. Analytic continuation of the coupled channel $\pi\Sigma-\bar{K}N$ amplitude involves four 
different Riemann sheets, each labelled by the sign of the imaginary parts of 
$(k_{\pi\Sigma},k_{\bar{K}N})$, with $(+,+)$ denoting the physical sheet. Complex poles in the 
scattering amplitude correspond to vanishing eigenvalues in the right-hand side of 
Eq.~(\ref{eq:amplitude}), and are determined numerically. In the vicinity of a pole, the divergent 
part of the amplitude is
\begin{equation}
     t =  \frac{m_\pi}{  E_{\rm cm} -  E_{\rm pole}} \begin{pmatrix}
   c_{\pi\Sigma}^2 & c_{\pi\Sigma} \, c_{\bar{K}N} \\c_{\pi\Sigma} \, c_{ \bar{K}N}  &  c_{ \bar{K}N}^2
\end{pmatrix}\,+\,\dots,
\end{equation}
where the (complex) residues $c_{\pi\Sigma}$ and  $c_{\bar{K}N}$ denote the coupling of the resonance 
pole to each channel.

\begin{figure}
\centering
\includegraphics[width=\linewidth]{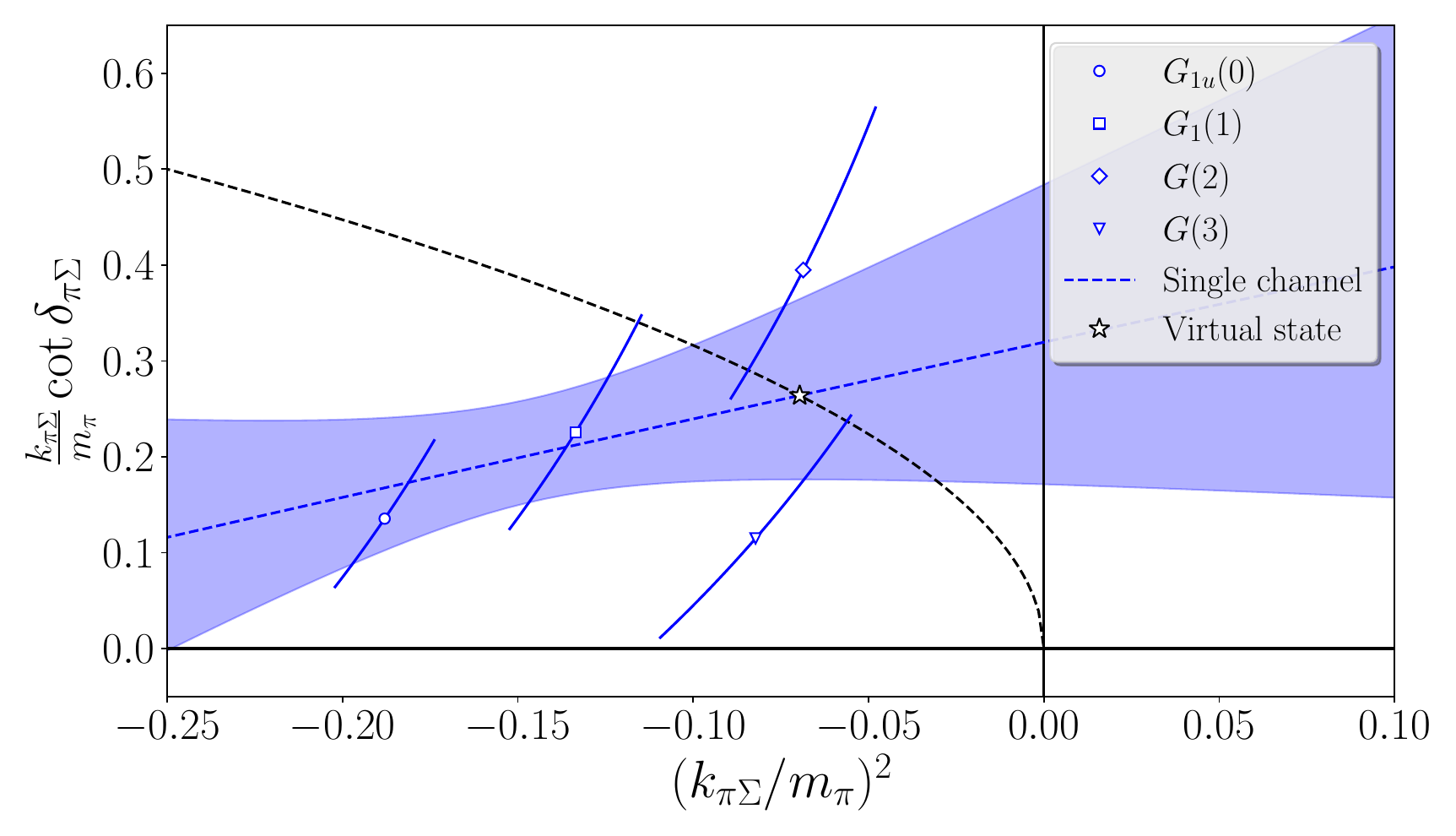}
\caption{The elastic $\pi\Sigma$ amplitude near threshold. The points are obtained from 
Eq.~(\ref{e:det}) using a single channel and $\ell_{\rm max} = 0$. The shaded band denotes a 
fit of the four levels shown to a two-parameter effective range expansion.  A pole on the real 
axis in the second Riemann sheet (a virtual bound state) occurs when $k_{\pi\Sigma}\cot 
\delta_{\pi\Sigma} - ik_{\pi\Sigma}=0$ below threshold. This is where the black dashed line 
intersects the fit.
\label{fig:swave} }
\end{figure}

Two poles are found on the  $(-,+)$ sheet,  which is the one closest to physical scattering in the 
region between the two thresholds. Their locations are
\begin{align}
\begin{split}
E_1 =& 1392(9)(2)(16) \text{ MeV}, \\
E_2 =&  [1455(13)(2)(17) - i 11.5(4.4)(4)(0.1)] \text{ MeV}, 
\end{split}
\label{eq:poles}
\end{align}
and their couplings
\begin{equation}
\left|\frac{c^{(1)}_{\pi\Sigma}}{c^{(1)}_{ \bar{K}N}}\right| = 1.9(4)(6), \qquad 
\left| \frac{c^{(2)}_{\pi\Sigma}}{c^{(2)}_{ \bar{K}N}}\right| = 0.53(9)(10).
\label{eq:couplings}
\end{equation}
The first uncertainty is statistical, the second accounts for parametrization dependence, and for 
the pole positions, the third comes from the uncertainty in the lattice spacing in 
Table~\ref{tab:comp_deets}. Two poles are present for all parametrizations of the $K$-matrix. 
The pole at $E_1$ is likely a virtual bound state, except in 0.5\% of bootstrap samples where 
it is located on the physical sheet and  thus a bound state, while the one at $E_2$ is a resonance. 
The first pole has a stronger coupling to the $\pi \Sigma$ channel, while for the second, the 
hierarchy is reversed, a pattern also predicted by chiral unitary models. Further confirmation of
the existence of the lower pole as a virtual bound
state comes from a single-channel analysis of the energy levels near the $\pi \Sigma$ threshold, 
as shown in Fig.~\ref{fig:swave}. 

{\it Conclusion.}---
This study of $\pi\Sigma-\bar{K}N$ scattering in the $\Lambda(1405)$ region is the first 
coupled-channel meson-baryon scattering amplitude determined from lattice QCD. Hermitian 
correlation matrices using both single-baryon and meson-baryon interpolating operators for 
a variety of different total momenta and irreducible representations were used. The analytic 
continuation of the amplitudes into the complex center-of-mass energy plane is stabilized by 
finite-volume energies just below the $\pi\Sigma$ and $\bar{K}N$ thresholds and clearly exhibits 
two poles. At a slightly heavier-than-physical pion mass of $m_{\pi} \approx 200\,{\rm MeV}$, 
the lower pole is a virtual bound state below the $\pi\Sigma$ threshold and the higher a 
resonance  just below the $\bar{K}N$ threshold. Due to our use of $m_\pi > m_\pi^{\rm phys}$, 
the real parts of the pole positions in Eq.~(\ref{eq:poles}) are somewhat larger than those 
determined at the physical point from experiment using chiral 
approaches~\cite{ParticleDataGroup:2022pth}, which lie within the ranges 
${\rm Re} \, E_1 = 1325-1380\,{\rm MeV}$ and ${\rm Re }\, E_2 = 1421-1434\,{\rm MeV}$. 
Importantly, this qualitative consistency supports the two-pole picture predicted by chiral 
symmetry and unitarity.  

Future work with this system includes moving to physical quark masses which requires the 
consideration of three particle effects, but this should not present a major problem in the 
region relevant for the $\Lambda(1405)$. Estimating residual finite-volume and lattice spacing 
effects are also planned. Studying this system along the quark-mass trajectory toward the 
SU(3)-symmetric point will also test the motion of the pole positions predicted by chiral 
effective theories.  Finally, this work opens the door to investigations of other baryon 
resonances, such as the $N(1535)$, $\Lambda(1670)$, $\Sigma(1620)$, and $\Xi(1620)$.

\vspace{4mm} 
\noindent 
{\bf Acknowledgements}:
We acknowledge helpful discussions with R.J.~Hudspith, A.~Jackura, M.~Mai, and M.~Lutz. We are also grateful 
to S.~Kuberski for providing re-weighting factors for the D200 ensemble, as computed according to 
Ref.~\cite{Kuberski:2023zky}. We thank our colleagues within the CLS consortium for sharing ensembles.
Computations were carried out on Frontera~\cite{frontera} at the Texas Advanced Computing Center (TACC), and 
at the National Energy Research Scientific Computing Center (NERSC), a U.S.~Department of Energy (DOE) Office 
of Science User Facility located at Lawrence Berkeley National Laboratory, operated under 
Contract No.~DE-AC02-05CH11231 using NERSC awards NP-ERCAP0005287, NP-ERCAP0010836 and NP-ERCAP0015497.
This work was supported in part by the U.S.~National Science Foundation under awards PHY-1913158 and 
PHY-2209167 (C.M., S.S.), the Faculty Early Career Development Program (CAREER) under award PHY-2047185 (A.N.)
and by the Graduate Research Fellowship Program under Grant No. DGE-2040435 (J.M.), the U.S.~Department of Energy, 
Office of Science, Office of Nuclear Physics, under grant contract numbers DE-SC0011090 and DE-SC0021006 (F.R.L.), 
DE-SC0012704 (A.D.H.), DE-AC02-05CH11231 (A.W.L.) and within the framework of Scientific Discovery through Advanced 
Computing (SciDAC) award ``Fundamental Nuclear Physics at the Exascale and Beyond'' (A.D.H.), the Mauricio and 
Carlota Botton Fellowship (F.R.L.), and the Heisenberg Programme of the Deutsche Forschungsgemeinschaft (DFG, 
German Research Foundation) project number 454605793 (D.M.).  NumPy~\cite{harris2020array}, 
matplotlib~\cite{Hunter:2007}, and the CHROMA software suite~\cite{Edwards:2004sx} were used for analysis, 
plotting, and correlator evaluation.

\bibliography{latticen}

\begin{thebibliography}{73}%
\makeatletter
\providecommand \@ifxundefined [1]{%
 \@ifx{#1\undefined}
}%
\providecommand \@ifnum [1]{%
 \ifnum #1\expandafter \@firstoftwo
 \else \expandafter \@secondoftwo
 \fi
}%
\providecommand \@ifx [1]{%
 \ifx #1\expandafter \@firstoftwo
 \else \expandafter \@secondoftwo
 \fi
}%
\providecommand \natexlab [1]{#1}%
\providecommand \enquote  [1]{``#1''}%
\providecommand \bibnamefont  [1]{#1}%
\providecommand \bibfnamefont [1]{#1}%
\providecommand \citenamefont [1]{#1}%
\providecommand \href@noop [0]{\@secondoftwo}%
\providecommand \href [0]{\begingroup \@sanitize@url \@href}%
\providecommand \@href[1]{\@@startlink{#1}\@@href}%
\providecommand \@@href[1]{\endgroup#1\@@endlink}%
\providecommand \@sanitize@url [0]{\catcode `\\12\catcode `\$12\catcode
  `\&12\catcode `\#12\catcode `\^12\catcode `\_12\catcode `\%12\relax}%
\providecommand \@@startlink[1]{}%
\providecommand \@@endlink[0]{}%
\providecommand \url  [0]{\begingroup\@sanitize@url \@url }%
\providecommand \@url [1]{\endgroup\@href {#1}{\urlprefix }}%
\providecommand \urlprefix  [0]{URL }%
\providecommand \Eprint [0]{\href }%
\providecommand \doibase [0]{https://doi.org/}%
\providecommand \selectlanguage [0]{\@gobble}%
\providecommand \bibinfo  [0]{\@secondoftwo}%
\providecommand \bibfield  [0]{\@secondoftwo}%
\providecommand \translation [1]{[#1]}%
\providecommand \BibitemOpen [0]{}%
\providecommand \bibitemStop [0]{}%
\providecommand \bibitemNoStop [0]{.\EOS\space}%
\providecommand \EOS [0]{\spacefactor3000\relax}%
\providecommand \BibitemShut  [1]{\csname bibitem#1\endcsname}%
\let\auto@bib@innerbib\@empty
\bibitem [{\citenamefont {Dalitz}\ and\ \citenamefont
  {Tuan}(1959)}]{Dalitz:1959dn}%
  \BibitemOpen
  \bibfield  {author} {\bibinfo {author} {\bibfnamefont {R.~H.}\ \bibnamefont
  {Dalitz}}\ and\ \bibinfo {author} {\bibfnamefont {S.~F.}\ \bibnamefont
  {Tuan}},\ }\bibfield  {title} {\bibinfo {title} {{Possible resonant state in
  pion-hyperon scattering}},\ }\href
  {https://doi.org/10.1103/PhysRevLett.2.425} {\bibfield  {journal} {\bibinfo
  {journal} {Phys. Rev. Lett.}\ }\textbf {\bibinfo {volume} {2}},\ \bibinfo
  {pages} {425} (\bibinfo {year} {1959})}\BibitemShut {NoStop}%
\bibitem [{\citenamefont {Dalitz}\ and\ \citenamefont
  {Tuan}(1960)}]{Dalitz:1960du}%
  \BibitemOpen
  \bibfield  {author} {\bibinfo {author} {\bibfnamefont {R.~H.}\ \bibnamefont
  {Dalitz}}\ and\ \bibinfo {author} {\bibfnamefont {S.~F.}\ \bibnamefont
  {Tuan}},\ }\bibfield  {title} {\bibinfo {title} {{The phenomenological
  representation of $\overline{K}$-nucleon scattering and reaction
  amplitudes}},\ }\href {https://doi.org/10.1016/0003-4916(60)90001-4}
  {\bibfield  {journal} {\bibinfo  {journal} {Annals Phys.}\ }\textbf {\bibinfo
  {volume} {10}},\ \bibinfo {pages} {307} (\bibinfo {year} {1960})}\BibitemShut
  {NoStop}%
\bibitem [{\citenamefont {Hyodo}\ and\ \citenamefont
  {Jido}(2012)}]{Hyodo:2011ur}%
  \BibitemOpen
  \bibfield  {author} {\bibinfo {author} {\bibfnamefont {T.}~\bibnamefont
  {Hyodo}}\ and\ \bibinfo {author} {\bibfnamefont {D.}~\bibnamefont {Jido}},\
  }\bibfield  {title} {\bibinfo {title} {{The nature of the $\Lambda(1405)$
  resonance in chiral dynamics}},\ }\href
  {https://doi.org/10.1016/j.ppnp.2011.07.002} {\bibfield  {journal} {\bibinfo
  {journal} {Prog. Part. Nucl. Phys.}\ }\textbf {\bibinfo {volume} {67}},\
  \bibinfo {pages} {55} (\bibinfo {year} {2012})},\ \Eprint
  {https://arxiv.org/abs/1104.4474} {arXiv:1104.4474 [nucl-th]} \BibitemShut
  {NoStop}%
\bibitem [{\citenamefont {Hyodo}\ and\ \citenamefont
  {Niiyama}(2021)}]{Hyodo:2020czb}%
  \BibitemOpen
  \bibfield  {author} {\bibinfo {author} {\bibfnamefont {T.}~\bibnamefont
  {Hyodo}}\ and\ \bibinfo {author} {\bibfnamefont {M.}~\bibnamefont
  {Niiyama}},\ }\bibfield  {title} {\bibinfo {title} {{QCD and the strange
  baryon spectrum}},\ }\href {https://doi.org/10.1016/j.ppnp.2021.103868}
  {\bibfield  {journal} {\bibinfo  {journal} {Prog. Part. Nucl. Phys.}\
  }\textbf {\bibinfo {volume} {120}},\ \bibinfo {pages} {103868} (\bibinfo
  {year} {2021})},\ \Eprint {https://arxiv.org/abs/2010.07592}
  {arXiv:2010.07592 [hep-ph]} \BibitemShut {NoStop}%
\bibitem [{\citenamefont {Workman}\ \emph {et~al.}(2022)\citenamefont {Workman}
  \emph {et~al.}}]{ParticleDataGroup:2022pth}%
  \BibitemOpen
  \bibfield  {author} {\bibinfo {author} {\bibfnamefont {R.~L.}\ \bibnamefont
  {Workman}} \emph {et~al.} (\bibinfo {collaboration} {Particle Data Group}),\
  }\bibfield  {title} {\bibinfo {title} {{Review of Particle Physics}},\ }\href
  {https://doi.org/10.1093/ptep/ptac097} {\bibfield  {journal} {\bibinfo
  {journal} {PTEP}\ }\textbf {\bibinfo {volume} {2022}},\ \bibinfo {pages}
  {083C01} (\bibinfo {year} {2022})}\BibitemShut {NoStop}%
\bibitem [{\citenamefont {Meissner}\ \emph {et~al.}(2004)\citenamefont
  {Meissner}, \citenamefont {Raha},\ and\ \citenamefont
  {Rusetsky}}]{Meissner:2004jr}%
  \BibitemOpen
  \bibfield  {author} {\bibinfo {author} {\bibfnamefont {U.~G.}\ \bibnamefont
  {Meissner}}, \bibinfo {author} {\bibfnamefont {U.}~\bibnamefont {Raha}},\
  and\ \bibinfo {author} {\bibfnamefont {A.}~\bibnamefont {Rusetsky}},\
  }\bibfield  {title} {\bibinfo {title} {{Spectrum and decays of kaonic
  hydrogen}},\ }\href {https://doi.org/10.1140/epjc/s2004-01859-4} {\bibfield
  {journal} {\bibinfo  {journal} {Eur. Phys. J. C}\ }\textbf {\bibinfo {volume}
  {35}},\ \bibinfo {pages} {349} (\bibinfo {year} {2004})},\ \Eprint
  {https://arxiv.org/abs/hep-ph/0402261} {arXiv:hep-ph/0402261} \BibitemShut
  {NoStop}%
\bibitem [{\citenamefont {Bazzi}\ \emph {et~al.}(2011)\citenamefont {Bazzi}
  \emph {et~al.}}]{SIDDHARTA:2011dsy}%
  \BibitemOpen
  \bibfield  {author} {\bibinfo {author} {\bibfnamefont {M.}~\bibnamefont
  {Bazzi}} \emph {et~al.} (\bibinfo {collaboration} {SIDDHARTA}),\ }\bibfield
  {title} {\bibinfo {title} {{A New Measurement of Kaonic Hydrogen X-rays}},\
  }\href {https://doi.org/10.1016/j.physletb.2011.09.011} {\bibfield  {journal}
  {\bibinfo  {journal} {Phys. Lett. B}\ }\textbf {\bibinfo {volume} {704}},\
  \bibinfo {pages} {113} (\bibinfo {year} {2011})},\ \Eprint
  {https://arxiv.org/abs/1105.3090} {arXiv:1105.3090 [nucl-ex]} \BibitemShut
  {NoStop}%
\bibitem [{\citenamefont {Moriya}\ \emph {et~al.}(2013)\citenamefont {Moriya}
  \emph {et~al.}}]{CLAS:2013rjt}%
  \BibitemOpen
  \bibfield  {author} {\bibinfo {author} {\bibfnamefont {K.}~\bibnamefont
  {Moriya}} \emph {et~al.} (\bibinfo {collaboration} {CLAS}),\ }\bibfield
  {title} {\bibinfo {title} {{Measurement of the
  \ensuremath{\Sigma}\ensuremath{\pi} photoproduction line shapes near the
  \ensuremath{\Lambda}(1405)}},\ }\href
  {https://doi.org/10.1103/PhysRevC.87.035206} {\bibfield  {journal} {\bibinfo
  {journal} {Phys. Rev. C}\ }\textbf {\bibinfo {volume} {87}},\ \bibinfo
  {pages} {035206} (\bibinfo {year} {2013})},\ \Eprint
  {https://arxiv.org/abs/1301.5000} {arXiv:1301.5000 [nucl-ex]} \BibitemShut
  {NoStop}%
\bibitem [{\citenamefont {Moriya}\ \emph {et~al.}(2014)\citenamefont {Moriya}
  \emph {et~al.}}]{PhysRevLett.112.082004}%
  \BibitemOpen
  \bibfield  {author} {\bibinfo {author} {\bibfnamefont {K.}~\bibnamefont
  {Moriya}} \emph {et~al.} (\bibinfo {collaboration} {CLAS Collaboration}),\
  }\bibfield  {title} {\bibinfo {title} {Spin and parity measurement of the
  $\mathrm{\ensuremath{\Lambda}}(1405)$ baryon},\ }\href
  {https://doi.org/10.1103/PhysRevLett.112.082004} {\bibfield  {journal}
  {\bibinfo  {journal} {Phys. Rev. Lett.}\ }\textbf {\bibinfo {volume} {112}},\
  \bibinfo {pages} {082004} (\bibinfo {year} {2014})}\BibitemShut {NoStop}%
\bibitem [{\citenamefont {Mai}\ and\ \citenamefont
  {Mei\ss{}ner}(2015)}]{Mai:2014xna}%
  \BibitemOpen
  \bibfield  {author} {\bibinfo {author} {\bibfnamefont {M.}~\bibnamefont
  {Mai}}\ and\ \bibinfo {author} {\bibfnamefont {U.-G.}\ \bibnamefont
  {Mei\ss{}ner}},\ }\bibfield  {title} {\bibinfo {title} {{Constraints on the
  chiral unitary $\bar KN$ amplitude from $\pi\Sigma K^+$ photoproduction
  data}},\ }\href {https://doi.org/10.1140/epja/i2015-15030-3} {\bibfield
  {journal} {\bibinfo  {journal} {Eur. Phys. J. A}\ }\textbf {\bibinfo {volume}
  {51}},\ \bibinfo {pages} {30} (\bibinfo {year} {2015})},\ \Eprint
  {https://arxiv.org/abs/1411.7884} {arXiv:1411.7884 [hep-ph]} \BibitemShut
  {NoStop}%
\bibitem [{\citenamefont {Roca}\ and\ \citenamefont
  {Oset}(2013)}]{Roca:2013cca}%
  \BibitemOpen
  \bibfield  {author} {\bibinfo {author} {\bibfnamefont {L.}~\bibnamefont
  {Roca}}\ and\ \bibinfo {author} {\bibfnamefont {E.}~\bibnamefont {Oset}},\
  }\bibfield  {title} {\bibinfo {title} {{Isospin 0 and 1 resonances from $\pi
  \Sigma$ photoproduction data}},\ }\href
  {https://doi.org/10.1103/PhysRevC.88.055206} {\bibfield  {journal} {\bibinfo
  {journal} {Phys. Rev. C}\ }\textbf {\bibinfo {volume} {88}},\ \bibinfo
  {pages} {055206} (\bibinfo {year} {2013})},\ \Eprint
  {https://arxiv.org/abs/1307.5752} {arXiv:1307.5752 [nucl-th]} \BibitemShut
  {NoStop}%
\bibitem [{\citenamefont {Scheluchin}\ \emph {et~al.}(2022)\citenamefont
  {Scheluchin} \emph {et~al.}}]{BGOOD:2021sog}%
  \BibitemOpen
  \bibfield  {author} {\bibinfo {author} {\bibfnamefont {G.}~\bibnamefont
  {Scheluchin}} \emph {et~al.} (\bibinfo {collaboration} {BGOOD}),\ }\bibfield
  {title} {\bibinfo {title} {{Photoproduction of $K^+\Lambda(1405)\rightarrow
  K^+\pi^0\Sigma^0$ extending to forward angles and low momentum transfer}},\
  }\href {https://doi.org/10.1016/j.physletb.2022.137375} {\bibfield  {journal}
  {\bibinfo  {journal} {Phys. Lett. B}\ }\textbf {\bibinfo {volume} {833}},\
  \bibinfo {pages} {137375} (\bibinfo {year} {2022})},\ \Eprint
  {https://arxiv.org/abs/2108.12235} {arXiv:2108.12235 [nucl-ex]} \BibitemShut
  {NoStop}%
\bibitem [{\citenamefont {Wickramaarachchi}\ \emph {et~al.}(2022)\citenamefont
  {Wickramaarachchi}, \citenamefont {Schumacher},\ and\ \citenamefont
  {Kalicy}}]{Wickramaarachchi:2022mhi}%
  \BibitemOpen
  \bibfield  {author} {\bibinfo {author} {\bibfnamefont {N.}~\bibnamefont
  {Wickramaarachchi}}, \bibinfo {author} {\bibfnamefont {R.~A.}\ \bibnamefont
  {Schumacher}},\ and\ \bibinfo {author} {\bibfnamefont {G.}~\bibnamefont
  {Kalicy}} (\bibinfo {collaboration} {GlueX}),\ }\bibfield  {title} {\bibinfo
  {title} {{Decay of the \ensuremath{\Lambda}(1405) hyperon to
  \ensuremath{\Sigma^0}\ensuremath{\pi^0} measured at GlueX}},\ }\href
  {https://doi.org/10.1051/epjconf/202227107005} {\bibfield  {journal}
  {\bibinfo  {journal} {EPJ Web Conf.}\ }\textbf {\bibinfo {volume} {271}},\
  \bibinfo {pages} {07005} (\bibinfo {year} {2022})},\ \Eprint
  {https://arxiv.org/abs/2209.06230} {arXiv:2209.06230 [nucl-ex]} \BibitemShut
  {NoStop}%
\bibitem [{\citenamefont {Acharya}\ \emph {et~al.}(2023)\citenamefont {Acharya}
  \emph {et~al.}}]{ALICE:2022yyh}%
  \BibitemOpen
  \bibfield  {author} {\bibinfo {author} {\bibfnamefont {S.}~\bibnamefont
  {Acharya}} \emph {et~al.} (\bibinfo {collaboration} {ALICE}),\ }\bibfield
  {title} {\bibinfo {title} {{Constraining the $\overline{K}\,N$ coupled
  channel dynamics using femtoscopic correlations at the LHC}},\ }\href
  {https://doi.org/10.1140/epjc/s10052-023-11476-0} {\bibfield  {journal}
  {\bibinfo  {journal} {Eur. Phys. J. C}\ }\textbf {\bibinfo {volume} {83}},\
  \bibinfo {pages} {340} (\bibinfo {year} {2023})},\ \Eprint
  {https://arxiv.org/abs/2205.15176} {arXiv:2205.15176 [nucl-ex]} \BibitemShut
  {NoStop}%
\bibitem [{\citenamefont {Aikawa}\ \emph {et~al.}(2023)\citenamefont {Aikawa}
  \emph {et~al.}}]{J-PARCE31:2022plu}%
  \BibitemOpen
  \bibfield  {author} {\bibinfo {author} {\bibfnamefont {S.}~\bibnamefont
  {Aikawa}} \emph {et~al.} (\bibinfo {collaboration} {J-PARC E31}),\ }\bibfield
   {title} {\bibinfo {title} {{Pole position of \ensuremath{\Lambda}(1405)
  measured in d(K\ensuremath{^-},n)\ensuremath{\pi}\ensuremath{\Sigma}
  reactions}},\ }\href {https://doi.org/10.1016/j.physletb.2022.137637}
  {\bibfield  {journal} {\bibinfo  {journal} {Phys. Lett. B}\ }\textbf
  {\bibinfo {volume} {837}},\ \bibinfo {pages} {137637} (\bibinfo {year}
  {2023})},\ \Eprint {https://arxiv.org/abs/2209.08254} {arXiv:2209.08254
  [nucl-ex]} \BibitemShut {NoStop}%
\bibitem [{\citenamefont {Anisovich}\ \emph {et~al.}(2020)\citenamefont
  {Anisovich}, \citenamefont {Sarantsev}, \citenamefont {Nikonov},
  \citenamefont {Burkert}, \citenamefont {Schumacher}, \citenamefont {Thoma},\
  and\ \citenamefont {Klempt}}]{Anisovich:2020lec}%
  \BibitemOpen
  \bibfield  {author} {\bibinfo {author} {\bibfnamefont {A.~V.}\ \bibnamefont
  {Anisovich}}, \bibinfo {author} {\bibfnamefont {A.~V.}\ \bibnamefont
  {Sarantsev}}, \bibinfo {author} {\bibfnamefont {V.~A.}\ \bibnamefont
  {Nikonov}}, \bibinfo {author} {\bibfnamefont {V.}~\bibnamefont {Burkert}},
  \bibinfo {author} {\bibfnamefont {R.~A.}\ \bibnamefont {Schumacher}},
  \bibinfo {author} {\bibfnamefont {U.}~\bibnamefont {Thoma}},\ and\ \bibinfo
  {author} {\bibfnamefont {E.}~\bibnamefont {Klempt}},\ }\bibfield  {title}
  {\bibinfo {title} {{Hyperon III: $K^{-}p - \pi \Sigma $ coupled-channel
  dynamics in the $\Lambda (1405)$ mass region}},\ }\href
  {https://doi.org/10.1140/epja/s10050-020-00142-8} {\bibfield  {journal}
  {\bibinfo  {journal} {Eur. Phys. J. A}\ }\textbf {\bibinfo {volume} {56}},\
  \bibinfo {pages} {139} (\bibinfo {year} {2020})}\BibitemShut {NoStop}%
\bibitem [{\citenamefont {Isgur}\ and\ \citenamefont
  {Karl}(1978)}]{Isgur:1978xj}%
  \BibitemOpen
  \bibfield  {author} {\bibinfo {author} {\bibfnamefont {N.}~\bibnamefont
  {Isgur}}\ and\ \bibinfo {author} {\bibfnamefont {G.}~\bibnamefont {Karl}},\
  }\bibfield  {title} {\bibinfo {title} {{P Wave Baryons in the Quark Model}},\
  }\href {https://doi.org/10.1103/PhysRevD.18.4187} {\bibfield  {journal}
  {\bibinfo  {journal} {Phys. Rev. D}\ }\textbf {\bibinfo {volume} {18}},\
  \bibinfo {pages} {4187} (\bibinfo {year} {1978})}\BibitemShut {NoStop}%
\bibitem [{\citenamefont {Kaiser}\ \emph {et~al.}(1995)\citenamefont {Kaiser},
  \citenamefont {Siegel},\ and\ \citenamefont {Weise}}]{Kaiser:1995eg}%
  \BibitemOpen
  \bibfield  {author} {\bibinfo {author} {\bibfnamefont {N.}~\bibnamefont
  {Kaiser}}, \bibinfo {author} {\bibfnamefont {P.~B.}\ \bibnamefont {Siegel}},\
  and\ \bibinfo {author} {\bibfnamefont {W.}~\bibnamefont {Weise}},\ }\bibfield
   {title} {\bibinfo {title} {{Chiral dynamics and the low-energy kaon -
  nucleon interaction}},\ }\href {https://doi.org/10.1016/0375-9474(95)00362-5}
  {\bibfield  {journal} {\bibinfo  {journal} {Nucl. Phys. A}\ }\textbf
  {\bibinfo {volume} {594}},\ \bibinfo {pages} {325} (\bibinfo {year}
  {1995})},\ \Eprint {https://arxiv.org/abs/nucl-th/9505043}
  {arXiv:nucl-th/9505043} \BibitemShut {NoStop}%
\bibitem [{\citenamefont {Oset}\ and\ \citenamefont
  {Ramos}(1998)}]{Oset:1997it}%
  \BibitemOpen
  \bibfield  {author} {\bibinfo {author} {\bibfnamefont {E.}~\bibnamefont
  {Oset}}\ and\ \bibinfo {author} {\bibfnamefont {A.}~\bibnamefont {Ramos}},\
  }\bibfield  {title} {\bibinfo {title} {{Nonperturbative chiral approach to
  $s$ wave $\overline{K} N$ interactions}},\ }\href
  {https://doi.org/10.1016/S0375-9474(98)00170-5} {\bibfield  {journal}
  {\bibinfo  {journal} {Nucl. Phys. A}\ }\textbf {\bibinfo {volume} {635}},\
  \bibinfo {pages} {99} (\bibinfo {year} {1998})},\ \Eprint
  {https://arxiv.org/abs/nucl-th/9711022} {arXiv:nucl-th/9711022} \BibitemShut
  {NoStop}%
\bibitem [{\citenamefont {Martinez~Torres}\ \emph {et~al.}(2012)\citenamefont
  {Martinez~Torres}, \citenamefont {Bayar}, \citenamefont {Jido},\ and\
  \citenamefont {Oset}}]{MartinezTorres:2012yi}%
  \BibitemOpen
  \bibfield  {author} {\bibinfo {author} {\bibfnamefont {A.}~\bibnamefont
  {Martinez~Torres}}, \bibinfo {author} {\bibfnamefont {M.}~\bibnamefont
  {Bayar}}, \bibinfo {author} {\bibfnamefont {D.}~\bibnamefont {Jido}},\ and\
  \bibinfo {author} {\bibfnamefont {E.}~\bibnamefont {Oset}},\ }\bibfield
  {title} {\bibinfo {title} {{Strategy to find the two $\Lambda(1405)$ states
  from lattice QCD simulations}},\ }\href
  {https://doi.org/10.1103/PhysRevC.86.055201} {\bibfield  {journal} {\bibinfo
  {journal} {Phys. Rev. C}\ }\textbf {\bibinfo {volume} {86}},\ \bibinfo
  {pages} {055201} (\bibinfo {year} {2012})},\ \Eprint
  {https://arxiv.org/abs/1202.4297} {arXiv:1202.4297 [hep-lat]} \BibitemShut
  {NoStop}%
\bibitem [{\citenamefont {Weinberg}(1966)}]{Weinberg:1966kf}%
  \BibitemOpen
  \bibfield  {author} {\bibinfo {author} {\bibfnamefont {S.}~\bibnamefont
  {Weinberg}},\ }\bibfield  {title} {\bibinfo {title} {{Pion scattering
  lengths}},\ }\href {https://doi.org/10.1103/PhysRevLett.17.616} {\bibfield
  {journal} {\bibinfo  {journal} {Phys. Rev. Lett.}\ }\textbf {\bibinfo
  {volume} {17}},\ \bibinfo {pages} {616} (\bibinfo {year} {1966})}\BibitemShut
  {NoStop}%
\bibitem [{\citenamefont {Tomozawa}(1966)}]{Tomozawa:1966jm}%
  \BibitemOpen
  \bibfield  {author} {\bibinfo {author} {\bibfnamefont {Y.}~\bibnamefont
  {Tomozawa}},\ }\bibfield  {title} {\bibinfo {title} {{Axial vector coupling
  renormalization and the meson baryon scattering lengths}},\ }\href
  {https://doi.org/10.1007/BF02857517} {\bibfield  {journal} {\bibinfo
  {journal} {Nuovo Cim. A}\ }\textbf {\bibinfo {volume} {46}},\ \bibinfo
  {pages} {707} (\bibinfo {year} {1966})}\BibitemShut {NoStop}%
\bibitem [{\citenamefont {Oller}\ and\ \citenamefont
  {Meissner}(2001)}]{Oller:2000fj}%
  \BibitemOpen
  \bibfield  {author} {\bibinfo {author} {\bibfnamefont {J.~A.}\ \bibnamefont
  {Oller}}\ and\ \bibinfo {author} {\bibfnamefont {U.~G.}\ \bibnamefont
  {Meissner}},\ }\bibfield  {title} {\bibinfo {title} {{Chiral dynamics in the
  presence of bound states: Kaon nucleon interactions revisited}},\ }\href
  {https://doi.org/10.1016/S0370-2693(01)00078-8} {\bibfield  {journal}
  {\bibinfo  {journal} {Phys. Lett. B}\ }\textbf {\bibinfo {volume} {500}},\
  \bibinfo {pages} {263} (\bibinfo {year} {2001})},\ \Eprint
  {https://arxiv.org/abs/hep-ph/0011146} {arXiv:hep-ph/0011146} \BibitemShut
  {NoStop}%
\bibitem [{\citenamefont {Mai}(2018)}]{Mai:2018rjx}%
  \BibitemOpen
  \bibfield  {author} {\bibinfo {author} {\bibfnamefont {M.}~\bibnamefont
  {Mai}},\ }\bibfield  {title} {\bibinfo {title} {{Status of the $\Lambda
  (1405)$}},\ }\href {https://doi.org/10.1007/s00601-018-1389-4} {\bibfield
  {journal} {\bibinfo  {journal} {Few Body Syst.}\ }\textbf {\bibinfo {volume}
  {59}},\ \bibinfo {pages} {61} (\bibinfo {year} {2018})}\BibitemShut {NoStop}%
\bibitem [{\citenamefont {Mai}(2021)}]{Mai:2020ltx}%
  \BibitemOpen
  \bibfield  {author} {\bibinfo {author} {\bibfnamefont {M.}~\bibnamefont
  {Mai}},\ }\bibfield  {title} {\bibinfo {title} {{Review of the ${\Lambda
  }$(1405) A curious case of a strangeness resonance}},\ }\href
  {https://doi.org/10.1140/epjs/s11734-021-00144-7} {\bibfield  {journal}
  {\bibinfo  {journal} {Eur. Phys. J. ST}\ }\textbf {\bibinfo {volume} {230}},\
  \bibinfo {pages} {1593} (\bibinfo {year} {2021})},\ \Eprint
  {https://arxiv.org/abs/2010.00056} {arXiv:2010.00056 [nucl-th]} \BibitemShut
  {NoStop}%
\bibitem [{\citenamefont {Mei\ss{}ner}(2020)}]{Meissner:2020khl}%
  \BibitemOpen
  \bibfield  {author} {\bibinfo {author} {\bibfnamefont {U.-G.}\ \bibnamefont
  {Mei\ss{}ner}},\ }\bibfield  {title} {\bibinfo {title} {{Two-pole structures
  in QCD: Facts, not fantasy!}},\ }\href {https://doi.org/10.3390/sym12060981}
  {\bibfield  {journal} {\bibinfo  {journal} {Symmetry}\ }\textbf {\bibinfo
  {volume} {12}},\ \bibinfo {pages} {981} (\bibinfo {year} {2020})},\ \Eprint
  {https://arxiv.org/abs/2005.06909} {arXiv:2005.06909 [hep-ph]} \BibitemShut
  {NoStop}%
\bibitem [{\citenamefont {Ezoe}\ and\ \citenamefont
  {Hosaka}(2020)}]{Ezoe:2020piq}%
  \BibitemOpen
  \bibfield  {author} {\bibinfo {author} {\bibfnamefont {T.}~\bibnamefont
  {Ezoe}}\ and\ \bibinfo {author} {\bibfnamefont {A.}~\bibnamefont {Hosaka}},\
  }\bibfield  {title} {\bibinfo {title} {{$\Lambda$(1405) as a $\overline{K}N$
  Feshbach resonance in the Skyrme model}},\ }\href
  {https://doi.org/10.1103/PhysRevD.102.014046} {\bibfield  {journal} {\bibinfo
   {journal} {Phys. Rev. D}\ }\textbf {\bibinfo {volume} {102}},\ \bibinfo
  {pages} {014046} (\bibinfo {year} {2020})},\ \Eprint
  {https://arxiv.org/abs/2006.03788} {arXiv:2006.03788 [hep-ph]} \BibitemShut
  {NoStop}%
\bibitem [{\citenamefont {Myint}\ \emph {et~al.}(2018)\citenamefont {Myint},
  \citenamefont {Akaishi}, \citenamefont {Hassanvand},\ and\ \citenamefont
  {Yamazaki}}]{Myint:2018ypc}%
  \BibitemOpen
  \bibfield  {author} {\bibinfo {author} {\bibfnamefont {K.~S.}\ \bibnamefont
  {Myint}}, \bibinfo {author} {\bibfnamefont {Y.}~\bibnamefont {Akaishi}},
  \bibinfo {author} {\bibfnamefont {M.}~\bibnamefont {Hassanvand}},\ and\
  \bibinfo {author} {\bibfnamefont {T.}~\bibnamefont {Yamazaki}},\ }\bibfield
  {title} {\bibinfo {title} {{Single-pole Nature of the Detectable
  Lambda(1405)}},\ }\href {https://doi.org/10.1093/ptep/pty075} {\bibfield
  {journal} {\bibinfo  {journal} {PTEP}\ }\textbf {\bibinfo {volume} {2018}},\
  \bibinfo {pages} {073D01} (\bibinfo {year} {2018})},\ \Eprint
  {https://arxiv.org/abs/1804.08240} {arXiv:1804.08240 [nucl-th]} \BibitemShut
  {NoStop}%
\bibitem [{\citenamefont {Miyahara}\ \emph {et~al.}(2018)\citenamefont
  {Miyahara}, \citenamefont {Hyodo},\ and\ \citenamefont
  {Weise}}]{Miyahara:2018onh}%
  \BibitemOpen
  \bibfield  {author} {\bibinfo {author} {\bibfnamefont {K.}~\bibnamefont
  {Miyahara}}, \bibinfo {author} {\bibfnamefont {T.}~\bibnamefont {Hyodo}},\
  and\ \bibinfo {author} {\bibfnamefont {W.}~\bibnamefont {Weise}},\ }\bibfield
   {title} {\bibinfo {title} {{Construction of a local $\bar K N-\pi \Sigma-\pi
  \Lambda$ potential and composition of the $\Lambda(1405)$}},\ }\href
  {https://doi.org/10.1103/PhysRevC.98.025201} {\bibfield  {journal} {\bibinfo
  {journal} {Phys. Rev. C}\ }\textbf {\bibinfo {volume} {98}},\ \bibinfo
  {pages} {025201} (\bibinfo {year} {2018})},\ \Eprint
  {https://arxiv.org/abs/1804.08269} {arXiv:1804.08269 [nucl-th]} \BibitemShut
  {NoStop}%
\bibitem [{\citenamefont {Miyahara}\ and\ \citenamefont
  {Hyodo}(2018)}]{Miyahara:2018lud}%
  \BibitemOpen
  \bibfield  {author} {\bibinfo {author} {\bibfnamefont {K.}~\bibnamefont
  {Miyahara}}\ and\ \bibinfo {author} {\bibfnamefont {T.}~\bibnamefont
  {Hyodo}},\ }\bibfield  {title} {\bibinfo {title} {{Theoretical study of
  $\Lambda(1405)$ resonance in $\Xi_b^0 \to D^0 (\pi \Sigma)$ decay}},\ }\href
  {https://doi.org/10.1103/PhysRevC.98.025202} {\bibfield  {journal} {\bibinfo
  {journal} {Phys. Rev. C}\ }\textbf {\bibinfo {volume} {98}},\ \bibinfo
  {pages} {025202} (\bibinfo {year} {2018})},\ \Eprint
  {https://arxiv.org/abs/1803.05572} {arXiv:1803.05572 [nucl-th]} \BibitemShut
  {NoStop}%
\bibitem [{\citenamefont {Azizi}\ \emph {et~al.}(2023)\citenamefont {Azizi},
  \citenamefont {Sarac},\ and\ \citenamefont {Sundu}}]{Azizi:2023tmw}%
  \BibitemOpen
  \bibfield  {author} {\bibinfo {author} {\bibfnamefont {K.}~\bibnamefont
  {Azizi}}, \bibinfo {author} {\bibfnamefont {Y.}~\bibnamefont {Sarac}},\ and\
  \bibinfo {author} {\bibfnamefont {H.}~\bibnamefont {Sundu}},\ }\bibfield
  {title} {\bibinfo {title} {{Investigation of $\Lambda(1405)$ as a molecular
  pentaquark state}},\ }\href@noop {} {\  (\bibinfo {year} {2023})},\ \Eprint
  {https://arxiv.org/abs/2306.07393} {arXiv:2306.07393 [hep-ph]} \BibitemShut
  {NoStop}%
\bibitem [{\citenamefont {Xie}\ \emph {et~al.}(2023)\citenamefont {Xie},
  \citenamefont {Lu}, \citenamefont {Geng},\ and\ \citenamefont
  {Zou}}]{Xie:2023cej}%
  \BibitemOpen
  \bibfield  {author} {\bibinfo {author} {\bibfnamefont {J.-M.}\ \bibnamefont
  {Xie}}, \bibinfo {author} {\bibfnamefont {J.-X.}\ \bibnamefont {Lu}},
  \bibinfo {author} {\bibfnamefont {L.-S.}\ \bibnamefont {Geng}},\ and\
  \bibinfo {author} {\bibfnamefont {B.-S.}\ \bibnamefont {Zou}},\ }\bibfield
  {title} {\bibinfo {title} {{Two-pole structures demystified: chiral dynamics
  at work}},\ }\href@noop {} {\bibfield  {journal} {\bibinfo  {journal}
  {(unpublished)}\ } (\bibinfo {year} {2023})},\ \Eprint
  {https://arxiv.org/abs/2307.11631} {arXiv:2307.11631 [hep-ph]} \BibitemShut
  {NoStop}%
\bibitem [{\citenamefont {Lu}\ \emph {et~al.}(2023)\citenamefont {Lu},
  \citenamefont {Geng}, \citenamefont {Doering},\ and\ \citenamefont
  {Mai}}]{Lu:2022hwm}%
  \BibitemOpen
  \bibfield  {author} {\bibinfo {author} {\bibfnamefont {J.-X.}\ \bibnamefont
  {Lu}}, \bibinfo {author} {\bibfnamefont {L.-S.}\ \bibnamefont {Geng}},
  \bibinfo {author} {\bibfnamefont {M.}~\bibnamefont {Doering}},\ and\ \bibinfo
  {author} {\bibfnamefont {M.}~\bibnamefont {Mai}},\ }\bibfield  {title}
  {\bibinfo {title} {{Cross-Channel Constraints on Resonant Antikaon-Nucleon
  Scattering}},\ }\href {https://doi.org/10.1103/PhysRevLett.130.071902}
  {\bibfield  {journal} {\bibinfo  {journal} {Phys. Rev. Lett.}\ }\textbf
  {\bibinfo {volume} {130}},\ \bibinfo {pages} {071902} (\bibinfo {year}
  {2023})},\ \Eprint {https://arxiv.org/abs/2209.02471} {arXiv:2209.02471
  [hep-ph]} \BibitemShut {NoStop}%
\bibitem [{\citenamefont {Hyodo}\ and\ \citenamefont
  {Weise}(2022)}]{Hyodo:2022xhp}%
  \BibitemOpen
  \bibfield  {author} {\bibinfo {author} {\bibfnamefont {T.}~\bibnamefont
  {Hyodo}}\ and\ \bibinfo {author} {\bibfnamefont {W.}~\bibnamefont {Weise}},\
  }\bibinfo {title} {{Theory of Kaon-Nuclear Systems}},\ in\ \href
  {https://doi.org/10.1007/978-981-15-8818-1_38-1} {\emph {\bibinfo {booktitle}
  {{Handbook of Nuclear Physics}}}},\ \bibinfo {editor} {edited by\ \bibinfo
  {editor} {\bibfnamefont {I.}~\bibnamefont {Tanihata}}, \bibinfo {editor}
  {\bibfnamefont {H.}~\bibnamefont {Toki}},\ and\ \bibinfo {editor}
  {\bibfnamefont {T.}~\bibnamefont {Kajino}}}\ (\bibinfo {year} {2022})\ pp.\
  \bibinfo {pages} {1--34},\ \Eprint {https://arxiv.org/abs/2202.06181}
  {arXiv:2202.06181 [nucl-th]} \BibitemShut {NoStop}%
\bibitem [{\citenamefont {Jido}\ \emph {et~al.}(2003)\citenamefont {Jido},
  \citenamefont {Oller}, \citenamefont {Oset}, \citenamefont {Ramos},\ and\
  \citenamefont {Meissner}}]{Jido:2003cb}%
  \BibitemOpen
  \bibfield  {author} {\bibinfo {author} {\bibfnamefont {D.}~\bibnamefont
  {Jido}}, \bibinfo {author} {\bibfnamefont {J.~A.}\ \bibnamefont {Oller}},
  \bibinfo {author} {\bibfnamefont {E.}~\bibnamefont {Oset}}, \bibinfo {author}
  {\bibfnamefont {A.}~\bibnamefont {Ramos}},\ and\ \bibinfo {author}
  {\bibfnamefont {U.~G.}\ \bibnamefont {Meissner}},\ }\bibfield  {title}
  {\bibinfo {title} {{Chiral dynamics of the two $\Lambda(1405)$ states}},\
  }\href {https://doi.org/10.1016/S0375-9474(03)01598-7} {\bibfield  {journal}
  {\bibinfo  {journal} {Nucl. Phys. A}\ }\textbf {\bibinfo {volume} {725}},\
  \bibinfo {pages} {181} (\bibinfo {year} {2003})},\ \Eprint
  {https://arxiv.org/abs/nucl-th/0303062} {arXiv:nucl-th/0303062} \BibitemShut
  {NoStop}%
\bibitem [{\citenamefont {Molina}\ and\ \citenamefont
  {D\"oring}(2016)}]{Molina:2015uqp}%
  \BibitemOpen
  \bibfield  {author} {\bibinfo {author} {\bibfnamefont {R.}~\bibnamefont
  {Molina}}\ and\ \bibinfo {author} {\bibfnamefont {M.}~\bibnamefont
  {D\"oring}},\ }\bibfield  {title} {\bibinfo {title} {{Pole structure of the
  $\Lambda$(1405) in a recent QCD simulation}},\ }\href
  {https://doi.org/10.1103/PhysRevD.94.079901} {\bibfield  {journal} {\bibinfo
  {journal} {Phys. Rev. D}\ }\textbf {\bibinfo {volume} {94}},\ \bibinfo
  {pages} {056010} (\bibinfo {year} {2016})},\ \bibinfo {note} {[Addendum:
  Phys.Rev.D 94, 079901 (2016)]},\ \Eprint {https://arxiv.org/abs/1512.05831}
  {arXiv:1512.05831 [hep-lat]} \BibitemShut {NoStop}%
\bibitem [{\citenamefont {{L\"{u}scher}}(1991)}]{Luscher:1990ux}%
  \BibitemOpen
  \bibfield  {author} {\bibinfo {author} {\bibfnamefont {M.}~\bibnamefont
  {{L\"{u}scher}}},\ }\bibfield  {title} {\bibinfo {title} {{Two particle
  states on a torus and their relation to the scattering matrix}},\ }\href
  {https://doi.org/10.1016/0550-3213(91)90366-6} {\bibfield  {journal}
  {\bibinfo  {journal} {Nucl. Phys.}\ }\textbf {\bibinfo {volume} {B354}},\
  \bibinfo {pages} {531} (\bibinfo {year} {1991})}\BibitemShut {NoStop}%
\bibitem [{\citenamefont {Rummukainen}\ and\ \citenamefont
  {Gottlieb}(1995)}]{Rummukainen:1995vs}%
  \BibitemOpen
  \bibfield  {author} {\bibinfo {author} {\bibfnamefont {K.}~\bibnamefont
  {Rummukainen}}\ and\ \bibinfo {author} {\bibfnamefont {S.~A.}\ \bibnamefont
  {Gottlieb}},\ }\bibfield  {title} {\bibinfo {title} {{Resonance scattering
  phase shifts on a nonrest frame lattice}},\ }\href
  {https://doi.org/10.1016/0550-3213(95)00313-H} {\bibfield  {journal}
  {\bibinfo  {journal} {Nucl. Phys. B}\ }\textbf {\bibinfo {volume} {450}},\
  \bibinfo {pages} {397} (\bibinfo {year} {1995})},\ \Eprint
  {https://arxiv.org/abs/hep-lat/9503028} {arXiv:hep-lat/9503028} \BibitemShut
  {NoStop}%
\bibitem [{\citenamefont {Kim}\ \emph {et~al.}(2005)\citenamefont {Kim},
  \citenamefont {Sachrajda},\ and\ \citenamefont {Sharpe}}]{Kim:2005gf}%
  \BibitemOpen
  \bibfield  {author} {\bibinfo {author} {\bibfnamefont {C.~H.}\ \bibnamefont
  {Kim}}, \bibinfo {author} {\bibfnamefont {C.~T.}\ \bibnamefont {Sachrajda}},\
  and\ \bibinfo {author} {\bibfnamefont {S.~R.}\ \bibnamefont {Sharpe}},\
  }\bibfield  {title} {\bibinfo {title} {{Finite-volume effects for two-hadron
  states in moving frames}},\ }\href
  {https://doi.org/10.1016/j.nuclphysb.2005.08.029} {\bibfield  {journal}
  {\bibinfo  {journal} {Nucl. Phys.}\ }\textbf {\bibinfo {volume} {B727}},\
  \bibinfo {pages} {218} (\bibinfo {year} {2005})},\ \Eprint
  {https://arxiv.org/abs/hep-lat/0507006} {arXiv:hep-lat/0507006 [hep-lat]}
  \BibitemShut {NoStop}%
\bibitem [{\citenamefont {He}\ \emph {et~al.}(2005)\citenamefont {He},
  \citenamefont {Feng},\ and\ \citenamefont {Liu}}]{He:2005ey}%
  \BibitemOpen
  \bibfield  {author} {\bibinfo {author} {\bibfnamefont {S.}~\bibnamefont
  {He}}, \bibinfo {author} {\bibfnamefont {X.}~\bibnamefont {Feng}},\ and\
  \bibinfo {author} {\bibfnamefont {C.}~\bibnamefont {Liu}},\ }\bibfield
  {title} {\bibinfo {title} {{Two particle states and the S-matrix elements in
  multi-channel scattering}},\ }\href
  {https://doi.org/10.1088/1126-6708/2005/07/011} {\bibfield  {journal}
  {\bibinfo  {journal} {JHEP}\ }\textbf {\bibinfo {volume} {07}},\ \bibinfo
  {pages} {011}},\ \Eprint {https://arxiv.org/abs/hep-lat/0504019}
  {arXiv:hep-lat/0504019 [hep-lat]} \BibitemShut {NoStop}%
\bibitem [{\citenamefont {Bernard}\ \emph {et~al.}(2011)\citenamefont
  {Bernard}, \citenamefont {Lage}, \citenamefont {Meissner},\ and\
  \citenamefont {Rusetsky}}]{Bernard:2010fp}%
  \BibitemOpen
  \bibfield  {author} {\bibinfo {author} {\bibfnamefont {V.}~\bibnamefont
  {Bernard}}, \bibinfo {author} {\bibfnamefont {M.}~\bibnamefont {Lage}},
  \bibinfo {author} {\bibfnamefont {U.~G.}\ \bibnamefont {Meissner}},\ and\
  \bibinfo {author} {\bibfnamefont {A.}~\bibnamefont {Rusetsky}},\ }\bibfield
  {title} {\bibinfo {title} {{Scalar mesons in a finite volume}},\ }\href
  {https://doi.org/10.1007/JHEP01(2011)019} {\bibfield  {journal} {\bibinfo
  {journal} {JHEP}\ }\textbf {\bibinfo {volume} {01}},\ \bibinfo {pages}
  {019}},\ \Eprint {https://arxiv.org/abs/1010.6018} {arXiv:1010.6018
  [hep-lat]} \BibitemShut {NoStop}%
\bibitem [{\citenamefont {{G\"{o}ckeler}}\ \emph {et~al.}(2012)\citenamefont
  {{G\"{o}ckeler}}, \citenamefont {Horsley}, \citenamefont {Lage},
  \citenamefont {{Mei{\ss}ner}}, \citenamefont {Rakow}, \citenamefont
  {Rusetsky}, \citenamefont {Schierholz},\ and\ \citenamefont
  {Zanotti}}]{Gockeler:2012yj}%
  \BibitemOpen
  \bibfield  {author} {\bibinfo {author} {\bibfnamefont {M.}~\bibnamefont
  {{G\"{o}ckeler}}}, \bibinfo {author} {\bibfnamefont {R.}~\bibnamefont
  {Horsley}}, \bibinfo {author} {\bibfnamefont {M.}~\bibnamefont {Lage}},
  \bibinfo {author} {\bibfnamefont {U.~G.}\ \bibnamefont {{Mei{\ss}ner}}},
  \bibinfo {author} {\bibfnamefont {P.~E.~L.}\ \bibnamefont {Rakow}}, \bibinfo
  {author} {\bibfnamefont {A.}~\bibnamefont {Rusetsky}}, \bibinfo {author}
  {\bibfnamefont {G.}~\bibnamefont {Schierholz}},\ and\ \bibinfo {author}
  {\bibfnamefont {J.~M.}\ \bibnamefont {Zanotti}},\ }\bibfield  {title}
  {\bibinfo {title} {{Scattering phases for meson and baryon resonances on
  general moving-frame lattices}},\ }\href
  {https://doi.org/10.1103/PhysRevD.86.094513} {\bibfield  {journal} {\bibinfo
  {journal} {Phys. Rev.}\ }\textbf {\bibinfo {volume} {D86}},\ \bibinfo {pages}
  {094513} (\bibinfo {year} {2012})},\ \Eprint
  {https://arxiv.org/abs/1206.4141} {arXiv:1206.4141 [hep-lat]} \BibitemShut
  {NoStop}%
\bibitem [{\citenamefont {Briceno}\ and\ \citenamefont
  {Davoudi}(2013)}]{Briceno:2012yi}%
  \BibitemOpen
  \bibfield  {author} {\bibinfo {author} {\bibfnamefont {R.~A.}\ \bibnamefont
  {Briceno}}\ and\ \bibinfo {author} {\bibfnamefont {Z.}~\bibnamefont
  {Davoudi}},\ }\bibfield  {title} {\bibinfo {title} {{Moving multichannel
  systems in a finite volume with application to proton-proton fusion}},\
  }\href {https://doi.org/10.1103/PhysRevD.88.094507} {\bibfield  {journal}
  {\bibinfo  {journal} {Phys. Rev.}\ }\textbf {\bibinfo {volume} {D88}},\
  \bibinfo {pages} {094507} (\bibinfo {year} {2013})},\ \Eprint
  {https://arxiv.org/abs/1204.1110} {arXiv:1204.1110 [hep-lat]} \BibitemShut
  {NoStop}%
\bibitem [{\citenamefont {Briceno}(2014)}]{Briceno:2014oea}%
  \BibitemOpen
  \bibfield  {author} {\bibinfo {author} {\bibfnamefont {R.~A.}\ \bibnamefont
  {Briceno}},\ }\bibfield  {title} {\bibinfo {title} {{Two-particle
  multichannel systems in a finite volume with arbitrary spin}},\ }\href
  {https://doi.org/10.1103/PhysRevD.89.074507} {\bibfield  {journal} {\bibinfo
  {journal} {Phys. Rev.}\ }\textbf {\bibinfo {volume} {D89}},\ \bibinfo {pages}
  {074507} (\bibinfo {year} {2014})},\ \Eprint
  {https://arxiv.org/abs/1401.3312} {arXiv:1401.3312 [hep-lat]} \BibitemShut
  {NoStop}%
\bibitem [{\citenamefont {Gubler}\ \emph {et~al.}(2016)\citenamefont {Gubler},
  \citenamefont {Takahashi},\ and\ \citenamefont {Oka}}]{Gubler:2016viv}%
  \BibitemOpen
  \bibfield  {author} {\bibinfo {author} {\bibfnamefont {P.}~\bibnamefont
  {Gubler}}, \bibinfo {author} {\bibfnamefont {T.~T.}\ \bibnamefont
  {Takahashi}},\ and\ \bibinfo {author} {\bibfnamefont {M.}~\bibnamefont
  {Oka}},\ }\bibfield  {title} {\bibinfo {title} {{Flavor structure of
  $\Lambda$ baryons from lattice QCD: From strange to charm quarks}},\ }\href
  {https://doi.org/10.1103/PhysRevD.94.114518} {\bibfield  {journal} {\bibinfo
  {journal} {Phys. Rev. D}\ }\textbf {\bibinfo {volume} {94}},\ \bibinfo
  {pages} {114518} (\bibinfo {year} {2016})},\ \Eprint
  {https://arxiv.org/abs/1609.01889} {arXiv:1609.01889 [hep-lat]} \BibitemShut
  {NoStop}%
\bibitem [{\citenamefont {Menadue}\ \emph {et~al.}(2012)\citenamefont
  {Menadue}, \citenamefont {Kamleh}, \citenamefont {Leinweber},\ and\
  \citenamefont {Mahbub}}]{Menadue:2011pd}%
  \BibitemOpen
  \bibfield  {author} {\bibinfo {author} {\bibfnamefont {B.~J.}\ \bibnamefont
  {Menadue}}, \bibinfo {author} {\bibfnamefont {W.}~\bibnamefont {Kamleh}},
  \bibinfo {author} {\bibfnamefont {D.~B.}\ \bibnamefont {Leinweber}},\ and\
  \bibinfo {author} {\bibfnamefont {M.~S.}\ \bibnamefont {Mahbub}},\ }\bibfield
   {title} {\bibinfo {title} {{Isolating the $\Lambda(1405)$ in Lattice QCD}},\
  }\href {https://doi.org/10.1103/PhysRevLett.108.112001} {\bibfield  {journal}
  {\bibinfo  {journal} {Phys. Rev. Lett.}\ }\textbf {\bibinfo {volume} {108}},\
  \bibinfo {pages} {112001} (\bibinfo {year} {2012})},\ \Eprint
  {https://arxiv.org/abs/1109.6716} {arXiv:1109.6716 [hep-lat]} \BibitemShut
  {NoStop}%
\bibitem [{\citenamefont {Engel}\ \emph
  {et~al.}(2013{\natexlab{a}})\citenamefont {Engel}, \citenamefont {Lang},\
  and\ \citenamefont {Sch\"afer}}]{Engel:2012qp}%
  \BibitemOpen
  \bibfield  {author} {\bibinfo {author} {\bibfnamefont {G.~P.}\ \bibnamefont
  {Engel}}, \bibinfo {author} {\bibfnamefont {C.~B.}\ \bibnamefont {Lang}},\
  and\ \bibinfo {author} {\bibfnamefont {A.}~\bibnamefont {Sch\"afer}}
  (\bibinfo {collaboration} {BGR (Bern-Graz-Regensburg)}),\ }\bibfield  {title}
  {\bibinfo {title} {{Low-lying $\Lambda$ baryons from the lattice}},\ }\href
  {https://doi.org/10.1103/PhysRevD.87.034502} {\bibfield  {journal} {\bibinfo
  {journal} {Phys. Rev. D}\ }\textbf {\bibinfo {volume} {87}},\ \bibinfo
  {pages} {034502} (\bibinfo {year} {2013}{\natexlab{a}})},\ \Eprint
  {https://arxiv.org/abs/1212.2032} {arXiv:1212.2032 [hep-lat]} \BibitemShut
  {NoStop}%
\bibitem [{\citenamefont {Engel}\ \emph
  {et~al.}(2013{\natexlab{b}})\citenamefont {Engel}, \citenamefont {Lang},
  \citenamefont {Mohler},\ and\ \citenamefont {Sch\"afer}}]{Engel:2013ig}%
  \BibitemOpen
  \bibfield  {author} {\bibinfo {author} {\bibfnamefont {G.~P.}\ \bibnamefont
  {Engel}}, \bibinfo {author} {\bibfnamefont {C.~B.}\ \bibnamefont {Lang}},
  \bibinfo {author} {\bibfnamefont {D.}~\bibnamefont {Mohler}},\ and\ \bibinfo
  {author} {\bibfnamefont {A.}~\bibnamefont {Sch\"afer}} (\bibinfo
  {collaboration} {BGR}),\ }\bibfield  {title} {\bibinfo {title} {{QCD with Two
  Light Dynamical Chirally Improved Quarks: Baryons}},\ }\href
  {https://doi.org/10.1103/PhysRevD.87.074504} {\bibfield  {journal} {\bibinfo
  {journal} {Phys. Rev. D}\ }\textbf {\bibinfo {volume} {87}},\ \bibinfo
  {pages} {074504} (\bibinfo {year} {2013}{\natexlab{b}})},\ \Eprint
  {https://arxiv.org/abs/1301.4318} {arXiv:1301.4318 [hep-lat]} \BibitemShut
  {NoStop}%
\bibitem [{\citenamefont {Nemoto}\ \emph {et~al.}(2003)\citenamefont {Nemoto},
  \citenamefont {Nakajima}, \citenamefont {Matsufuru},\ and\ \citenamefont
  {Suganuma}}]{Nemoto:2003ft}%
  \BibitemOpen
  \bibfield  {author} {\bibinfo {author} {\bibfnamefont {Y.}~\bibnamefont
  {Nemoto}}, \bibinfo {author} {\bibfnamefont {N.}~\bibnamefont {Nakajima}},
  \bibinfo {author} {\bibfnamefont {H.}~\bibnamefont {Matsufuru}},\ and\
  \bibinfo {author} {\bibfnamefont {H.}~\bibnamefont {Suganuma}},\ }\bibfield
  {title} {\bibinfo {title} {{Negative parity baryons in quenched anisotropic
  lattice QCD}},\ }\href {https://doi.org/10.1103/PhysRevD.68.094505}
  {\bibfield  {journal} {\bibinfo  {journal} {Phys. Rev. D}\ }\textbf {\bibinfo
  {volume} {68}},\ \bibinfo {pages} {094505} (\bibinfo {year} {2003})},\
  \Eprint {https://arxiv.org/abs/hep-lat/0302013} {arXiv:hep-lat/0302013}
  \BibitemShut {NoStop}%
\bibitem [{\citenamefont {Burch}\ \emph {et~al.}(2006)\citenamefont {Burch},
  \citenamefont {Gattringer}, \citenamefont {Glozman}, \citenamefont {Hagen},
  \citenamefont {Hierl}, \citenamefont {Lang},\ and\ \citenamefont
  {Schafer}}]{Burch:2006cc}%
  \BibitemOpen
  \bibfield  {author} {\bibinfo {author} {\bibfnamefont {T.}~\bibnamefont
  {Burch}}, \bibinfo {author} {\bibfnamefont {C.}~\bibnamefont {Gattringer}},
  \bibinfo {author} {\bibfnamefont {L.~Y.}\ \bibnamefont {Glozman}}, \bibinfo
  {author} {\bibfnamefont {C.}~\bibnamefont {Hagen}}, \bibinfo {author}
  {\bibfnamefont {D.}~\bibnamefont {Hierl}}, \bibinfo {author} {\bibfnamefont
  {C.~B.}\ \bibnamefont {Lang}},\ and\ \bibinfo {author} {\bibfnamefont
  {A.}~\bibnamefont {Schafer}},\ }\bibfield  {title} {\bibinfo {title}
  {{Excited hadrons on the lattice: Baryons}},\ }\href
  {https://doi.org/10.1103/PhysRevD.74.014504} {\bibfield  {journal} {\bibinfo
  {journal} {Phys. Rev. D}\ }\textbf {\bibinfo {volume} {74}},\ \bibinfo
  {pages} {014504} (\bibinfo {year} {2006})},\ \Eprint
  {https://arxiv.org/abs/hep-lat/0604019} {arXiv:hep-lat/0604019} \BibitemShut
  {NoStop}%
\bibitem [{\citenamefont {Takahashi}\ and\ \citenamefont
  {Oka}(2010)}]{Takahashi:2009bu}%
  \BibitemOpen
  \bibfield  {author} {\bibinfo {author} {\bibfnamefont {T.~T.}\ \bibnamefont
  {Takahashi}}\ and\ \bibinfo {author} {\bibfnamefont {M.}~\bibnamefont
  {Oka}},\ }\bibfield  {title} {\bibinfo {title} {{Low-lying Lambda Baryons
  with spin 1/2 in Two-flavor Lattice QCD}},\ }\href
  {https://doi.org/10.1103/PhysRevD.81.034505} {\bibfield  {journal} {\bibinfo
  {journal} {Phys. Rev. D}\ }\textbf {\bibinfo {volume} {81}},\ \bibinfo
  {pages} {034505} (\bibinfo {year} {2010})},\ \Eprint
  {https://arxiv.org/abs/0910.0686} {arXiv:0910.0686 [hep-lat]} \BibitemShut
  {NoStop}%
\bibitem [{\citenamefont {Meinel}\ and\ \citenamefont
  {Rendon}(2022)}]{Meinel:2021grq}%
  \BibitemOpen
  \bibfield  {author} {\bibinfo {author} {\bibfnamefont {S.}~\bibnamefont
  {Meinel}}\ and\ \bibinfo {author} {\bibfnamefont {G.}~\bibnamefont
  {Rendon}},\ }\bibfield  {title} {\bibinfo {title} {{Charm-baryon semileptonic
  decays and the strange \ensuremath{\Lambda}* resonances: New insights from
  lattice QCD}},\ }\href {https://doi.org/10.1103/PhysRevD.105.L051505}
  {\bibfield  {journal} {\bibinfo  {journal} {Phys. Rev. D}\ }\textbf {\bibinfo
  {volume} {105}},\ \bibinfo {pages} {L051505} (\bibinfo {year} {2022})},\
  \Eprint {https://arxiv.org/abs/2107.13084} {arXiv:2107.13084 [hep-ph]}
  \BibitemShut {NoStop}%
\bibitem [{\citenamefont {Hall}\ \emph {et~al.}(2015)\citenamefont {Hall},
  \citenamefont {Kamleh}, \citenamefont {Leinweber}, \citenamefont {Menadue},
  \citenamefont {Owen}, \citenamefont {Thomas},\ and\ \citenamefont
  {Young}}]{Hall:2014uca}%
  \BibitemOpen
  \bibfield  {author} {\bibinfo {author} {\bibfnamefont {J.~M.~M.}\
  \bibnamefont {Hall}}, \bibinfo {author} {\bibfnamefont {W.}~\bibnamefont
  {Kamleh}}, \bibinfo {author} {\bibfnamefont {D.~B.}\ \bibnamefont
  {Leinweber}}, \bibinfo {author} {\bibfnamefont {B.~J.}\ \bibnamefont
  {Menadue}}, \bibinfo {author} {\bibfnamefont {B.~J.}\ \bibnamefont {Owen}},
  \bibinfo {author} {\bibfnamefont {A.~W.}\ \bibnamefont {Thomas}},\ and\
  \bibinfo {author} {\bibfnamefont {R.~D.}\ \bibnamefont {Young}},\ }\bibfield
  {title} {\bibinfo {title} {{Lattice QCD Evidence that the
  \ensuremath{\Lambda}(1405) Resonance is an Antikaon-Nucleon Molecule}},\
  }\href {https://doi.org/10.1103/PhysRevLett.114.132002} {\bibfield  {journal}
  {\bibinfo  {journal} {Phys. Rev. Lett.}\ }\textbf {\bibinfo {volume} {114}},\
  \bibinfo {pages} {132002} (\bibinfo {year} {2015})},\ \Eprint
  {https://arxiv.org/abs/1411.3402} {arXiv:1411.3402 [hep-lat]} \BibitemShut
  {NoStop}%
\bibitem [{\citenamefont {Fukugita}\ \emph {et~al.}(1995)\citenamefont
  {Fukugita}, \citenamefont {Kuramashi}, \citenamefont {Okawa}, \citenamefont
  {Mino},\ and\ \citenamefont {Ukawa}}]{Fukugita:1994ve}%
  \BibitemOpen
  \bibfield  {author} {\bibinfo {author} {\bibfnamefont {M.}~\bibnamefont
  {Fukugita}}, \bibinfo {author} {\bibfnamefont {Y.}~\bibnamefont {Kuramashi}},
  \bibinfo {author} {\bibfnamefont {M.}~\bibnamefont {Okawa}}, \bibinfo
  {author} {\bibfnamefont {H.}~\bibnamefont {Mino}},\ and\ \bibinfo {author}
  {\bibfnamefont {A.}~\bibnamefont {Ukawa}},\ }\bibfield  {title} {\bibinfo
  {title} {{Hadron scattering lengths in lattice QCD}},\ }\href
  {https://doi.org/10.1103/PhysRevD.52.3003} {\bibfield  {journal} {\bibinfo
  {journal} {Phys. Rev.}\ }\textbf {\bibinfo {volume} {D52}},\ \bibinfo {pages}
  {3003} (\bibinfo {year} {1995})},\ \Eprint
  {https://arxiv.org/abs/hep-lat/9501024} {arXiv:hep-lat/9501024 [hep-lat]}
  \BibitemShut {NoStop}%
\bibitem [{\citenamefont {Detmold}\ and\ \citenamefont
  {Nicholson}(2016)}]{Detmold:2015qwf}%
  \BibitemOpen
  \bibfield  {author} {\bibinfo {author} {\bibfnamefont {W.}~\bibnamefont
  {Detmold}}\ and\ \bibinfo {author} {\bibfnamefont {A.}~\bibnamefont
  {Nicholson}},\ }\bibfield  {title} {\bibinfo {title} {{Low energy scattering
  phase shifts for meson-baryon systems}},\ }\href
  {https://doi.org/10.1103/PhysRevD.93.114511} {\bibfield  {journal} {\bibinfo
  {journal} {Phys. Rev.}\ }\textbf {\bibinfo {volume} {D93}},\ \bibinfo {pages}
  {114511} (\bibinfo {year} {2016})},\ \Eprint
  {https://arxiv.org/abs/1511.02275} {arXiv:1511.02275 [hep-lat]} \BibitemShut
  {NoStop}%
\bibitem [{\citenamefont {Torok}\ \emph {et~al.}(2010)\citenamefont {Torok},
  \citenamefont {Beane}, \citenamefont {Detmold}, \citenamefont {Luu},
  \citenamefont {Orginos}, \citenamefont {Parreno}, \citenamefont {Savage},\
  and\ \citenamefont {Walker-Loud}}]{Torok:2009dg}%
  \BibitemOpen
  \bibfield  {author} {\bibinfo {author} {\bibfnamefont {A.}~\bibnamefont
  {Torok}}, \bibinfo {author} {\bibfnamefont {S.~R.}\ \bibnamefont {Beane}},
  \bibinfo {author} {\bibfnamefont {W.}~\bibnamefont {Detmold}}, \bibinfo
  {author} {\bibfnamefont {T.~C.}\ \bibnamefont {Luu}}, \bibinfo {author}
  {\bibfnamefont {K.}~\bibnamefont {Orginos}}, \bibinfo {author} {\bibfnamefont
  {A.}~\bibnamefont {Parreno}}, \bibinfo {author} {\bibfnamefont {M.~J.}\
  \bibnamefont {Savage}},\ and\ \bibinfo {author} {\bibfnamefont
  {A.}~\bibnamefont {Walker-Loud}},\ }\bibfield  {title} {\bibinfo {title}
  {{Meson-Baryon Scattering Lengths from Mixed-Action Lattice QCD}},\ }\href
  {https://doi.org/10.1103/PhysRevD.81.074506} {\bibfield  {journal} {\bibinfo
  {journal} {Phys. Rev.}\ }\textbf {\bibinfo {volume} {D81}},\ \bibinfo {pages}
  {074506} (\bibinfo {year} {2010})},\ \Eprint
  {https://arxiv.org/abs/0907.1913} {arXiv:0907.1913 [hep-lat]} \BibitemShut
  {NoStop}%
\bibitem [{\citenamefont {Meng}\ \emph {et~al.}(2004)\citenamefont {Meng},
  \citenamefont {Miao}, \citenamefont {Du},\ and\ \citenamefont
  {Liu}}]{Meng:2003gm}%
  \BibitemOpen
  \bibfield  {author} {\bibinfo {author} {\bibfnamefont {G.-W.}\ \bibnamefont
  {Meng}}, \bibinfo {author} {\bibfnamefont {C.}~\bibnamefont {Miao}}, \bibinfo
  {author} {\bibfnamefont {X.-N.}\ \bibnamefont {Du}},\ and\ \bibinfo {author}
  {\bibfnamefont {C.}~\bibnamefont {Liu}},\ }\bibfield  {title} {\bibinfo
  {title} {{Lattice study on kaon nucleon scattering length in the $I = 1$
  channel}},\ }\href {https://doi.org/10.1142/S0217751X04019627} {\bibfield
  {journal} {\bibinfo  {journal} {Int. J. Mod. Phys. A}\ }\textbf {\bibinfo
  {volume} {19}},\ \bibinfo {pages} {4401} (\bibinfo {year} {2004})},\ \Eprint
  {https://arxiv.org/abs/hep-lat/0309048} {arXiv:hep-lat/0309048} \BibitemShut
  {NoStop}%
\bibitem [{\citenamefont {Bulava}\ \emph {et~al.}(2024)\citenamefont {Bulava},
  \citenamefont {Cid-Mora}, \citenamefont {Hanlon}, \citenamefont {H{\"{o}}rz},
  \citenamefont {Mohler}, \citenamefont {Morningstar}, \citenamefont {Moscoso},
  \citenamefont {Nicholson}, \citenamefont {Romero-L{\'{o}}pez}, \citenamefont
  {Skinner},\ and\ \citenamefont {Walker-Loud}}]{BaryonScatteringBaSc:2023ori}%
  \BibitemOpen
  \bibfield  {author} {\bibinfo {author} {\bibfnamefont {J.}~\bibnamefont
  {Bulava}}, \bibinfo {author} {\bibfnamefont {B.}~\bibnamefont {Cid-Mora}},
  \bibinfo {author} {\bibfnamefont {A.~D.}\ \bibnamefont {Hanlon}}, \bibinfo
  {author} {\bibfnamefont {B.}~\bibnamefont {H{\"{o}}rz}}, \bibinfo {author}
  {\bibfnamefont {D.}~\bibnamefont {Mohler}}, \bibinfo {author} {\bibfnamefont
  {C.}~\bibnamefont {Morningstar}}, \bibinfo {author} {\bibfnamefont
  {J.}~\bibnamefont {Moscoso}}, \bibinfo {author} {\bibfnamefont
  {A.}~\bibnamefont {Nicholson}}, \bibinfo {author} {\bibfnamefont
  {F.}~\bibnamefont {Romero-L{\'{o}}pez}}, \bibinfo {author} {\bibfnamefont
  {S.}~\bibnamefont {Skinner}},\ and\ \bibinfo {author} {\bibfnamefont
  {A.}~\bibnamefont {Walker-Loud}} (\bibinfo {collaboration} {Baryon Scattering
  (BaSc)}),\ }\bibfield  {title} {\bibinfo {title} {{Lattice QCD study of
  $\pi\Sigma-\bar{K}N$ scattering and the $\Lambda(1405)$ resonance}},\ }\href
  {https://doi.org/10.1103/PhysRevD.109.014511} {\bibfield  {journal} {\bibinfo
   {journal} {Phys. Rev. D}\ }\textbf {\bibinfo {volume} {109}},\ \bibinfo
  {pages} {014511} (\bibinfo {year} {2024})},\ \Eprint
  {https://arxiv.org/abs/2307.13471} {arXiv:2307.13471 [hep-lat]} \BibitemShut
  {NoStop}%
\bibitem [{\citenamefont {Bruno}\ \emph {et~al.}(2015)\citenamefont {Bruno}
  \emph {et~al.}}]{Bruno:2014jqa}%
  \BibitemOpen
  \bibfield  {author} {\bibinfo {author} {\bibfnamefont {M.}~\bibnamefont
  {Bruno}} \emph {et~al.},\ }\bibfield  {title} {\bibinfo {title} {{Simulation
  of QCD with N$_{f} =$ 2 $+$ 1 flavors of non-perturbatively improved Wilson
  fermions}},\ }\href {https://doi.org/10.1007/JHEP02(2015)043} {\bibfield
  {journal} {\bibinfo  {journal} {JHEP}\ }\textbf {\bibinfo {volume} {02}},\
  \bibinfo {pages} {043}},\ \Eprint {https://arxiv.org/abs/1411.3982}
  {arXiv:1411.3982 [hep-lat]} \BibitemShut {NoStop}%
\bibitem [{\citenamefont {Bruno}\ \emph {et~al.}(2017)\citenamefont {Bruno},
  \citenamefont {Korzec},\ and\ \citenamefont {Schaefer}}]{Bruno:2016plf}%
  \BibitemOpen
  \bibfield  {author} {\bibinfo {author} {\bibfnamefont {M.}~\bibnamefont
  {Bruno}}, \bibinfo {author} {\bibfnamefont {T.}~\bibnamefont {Korzec}},\ and\
  \bibinfo {author} {\bibfnamefont {S.}~\bibnamefont {Schaefer}},\ }\bibfield
  {title} {\bibinfo {title} {{Setting the scale for the CLS $2 + 1$ flavor
  ensembles}},\ }\href {https://doi.org/10.1103/PhysRevD.95.074504} {\bibfield
  {journal} {\bibinfo  {journal} {Phys. Rev.}\ }\textbf {\bibinfo {volume}
  {D95}},\ \bibinfo {pages} {074504} (\bibinfo {year} {2017})},\ \Eprint
  {https://arxiv.org/abs/1608.08900} {arXiv:1608.08900 [hep-lat]} \BibitemShut
  {NoStop}%
\bibitem [{\citenamefont {Strassberger}\ \emph {et~al.}(2022)\citenamefont
  {Strassberger} \emph {et~al.}}]{Strassberger:2021tsu}%
  \BibitemOpen
  \bibfield  {author} {\bibinfo {author} {\bibfnamefont {B.}~\bibnamefont
  {Strassberger}} \emph {et~al.},\ }\bibfield  {title} {\bibinfo {title}
  {{Scale setting for CLS 2+1 simulations}},\ }\href
  {https://doi.org/10.22323/1.396.0135} {\bibfield  {journal} {\bibinfo
  {journal} {PoS}\ }\textbf {\bibinfo {volume} {LATTICE2021}},\ \bibinfo
  {pages} {135} (\bibinfo {year} {2022})},\ \Eprint
  {https://arxiv.org/abs/2112.06696} {arXiv:2112.06696 [hep-lat]} \BibitemShut
  {NoStop}%
\bibitem [{\citenamefont {Morningstar}\ \emph {et~al.}(2011)\citenamefont
  {Morningstar}, \citenamefont {Bulava}, \citenamefont {Foley}, \citenamefont
  {Juge}, \citenamefont {Lenkner}, \citenamefont {Peardon},\ and\ \citenamefont
  {Wong}}]{Morningstar:2011ka}%
  \BibitemOpen
  \bibfield  {author} {\bibinfo {author} {\bibfnamefont {C.}~\bibnamefont
  {Morningstar}}, \bibinfo {author} {\bibfnamefont {J.}~\bibnamefont {Bulava}},
  \bibinfo {author} {\bibfnamefont {J.}~\bibnamefont {Foley}}, \bibinfo
  {author} {\bibfnamefont {K.~J.}\ \bibnamefont {Juge}}, \bibinfo {author}
  {\bibfnamefont {D.}~\bibnamefont {Lenkner}}, \bibinfo {author} {\bibfnamefont
  {M.}~\bibnamefont {Peardon}},\ and\ \bibinfo {author} {\bibfnamefont {C.~H.}\
  \bibnamefont {Wong}},\ }\bibfield  {title} {\bibinfo {title} {{Improved
  stochastic estimation of quark propagation with Laplacian Heaviside smearing
  in lattice QCD}},\ }\href {https://doi.org/10.1103/PhysRevD.83.114505}
  {\bibfield  {journal} {\bibinfo  {journal} {Phys. Rev.}\ }\textbf {\bibinfo
  {volume} {D83}},\ \bibinfo {pages} {114505} (\bibinfo {year} {2011})},\
  \Eprint {https://arxiv.org/abs/1104.3870} {arXiv:1104.3870 [hep-lat]}
  \BibitemShut {NoStop}%
\bibitem [{\citenamefont {Bulava}\ \emph {et~al.}(2023)\citenamefont {Bulava},
  \citenamefont {Hanlon}, \citenamefont {H\"orz}, \citenamefont {Morningstar},
  \citenamefont {Nicholson}, \citenamefont {Romero-L\'opez}, \citenamefont
  {Skinner}, \citenamefont {Vranas},\ and\ \citenamefont
  {Walker-Loud}}]{Bulava:2022vpq}%
  \BibitemOpen
  \bibfield  {author} {\bibinfo {author} {\bibfnamefont {J.}~\bibnamefont
  {Bulava}}, \bibinfo {author} {\bibfnamefont {A.~D.}\ \bibnamefont {Hanlon}},
  \bibinfo {author} {\bibfnamefont {B.}~\bibnamefont {H\"orz}}, \bibinfo
  {author} {\bibfnamefont {C.}~\bibnamefont {Morningstar}}, \bibinfo {author}
  {\bibfnamefont {A.}~\bibnamefont {Nicholson}}, \bibinfo {author}
  {\bibfnamefont {F.}~\bibnamefont {Romero-L\'opez}}, \bibinfo {author}
  {\bibfnamefont {S.}~\bibnamefont {Skinner}}, \bibinfo {author} {\bibfnamefont
  {P.}~\bibnamefont {Vranas}},\ and\ \bibinfo {author} {\bibfnamefont
  {A.}~\bibnamefont {Walker-Loud}},\ }\bibfield  {title} {\bibinfo {title}
  {{Elastic nucleon-pion scattering at $m_\pi=200$~MeV from lattice QCD}},\
  }\href {https://doi.org/10.1016/j.nuclphysb.2023.116105} {\bibfield
  {journal} {\bibinfo  {journal} {Nucl. Phys. B}\ }\textbf {\bibinfo {volume}
  {987}},\ \bibinfo {pages} {116105} (\bibinfo {year} {2023})},\ \Eprint
  {https://arxiv.org/abs/2208.03867} {arXiv:2208.03867 [hep-lat]} \BibitemShut
  {NoStop}%
\bibitem [{\citenamefont {H\"orz}\ and\ \citenamefont
  {Hanlon}(2019)}]{Horz:2019rrn}%
  \BibitemOpen
  \bibfield  {author} {\bibinfo {author} {\bibfnamefont {B.}~\bibnamefont
  {H\"orz}}\ and\ \bibinfo {author} {\bibfnamefont {A.}~\bibnamefont
  {Hanlon}},\ }\bibfield  {title} {\bibinfo {title} {{Two- and three-pion
  finite-volume spectra at maximal isospin from lattice QCD}},\ }\href
  {https://doi.org/10.1103/PhysRevLett.123.142002} {\bibfield  {journal}
  {\bibinfo  {journal} {Phys. Rev. Lett.}\ }\textbf {\bibinfo {volume} {123}},\
  \bibinfo {pages} {142002} (\bibinfo {year} {2019})},\ \Eprint
  {https://arxiv.org/abs/1905.04277} {arXiv:1905.04277 [hep-lat]} \BibitemShut
  {NoStop}%
\bibitem [{\citenamefont {Morningstar}\ \emph {et~al.}(2017)\citenamefont
  {Morningstar}, \citenamefont {Bulava}, \citenamefont {Singha}, \citenamefont
  {Brett}, \citenamefont {Fallica}, \citenamefont {Hanlon},\ and\ \citenamefont
  {{H\"{o}rz}}}]{Morningstar:2017spu}%
  \BibitemOpen
  \bibfield  {author} {\bibinfo {author} {\bibfnamefont {C.}~\bibnamefont
  {Morningstar}}, \bibinfo {author} {\bibfnamefont {J.}~\bibnamefont {Bulava}},
  \bibinfo {author} {\bibfnamefont {B.}~\bibnamefont {Singha}}, \bibinfo
  {author} {\bibfnamefont {R.}~\bibnamefont {Brett}}, \bibinfo {author}
  {\bibfnamefont {J.}~\bibnamefont {Fallica}}, \bibinfo {author} {\bibfnamefont
  {A.}~\bibnamefont {Hanlon}},\ and\ \bibinfo {author} {\bibfnamefont
  {B.}~\bibnamefont {{H\"{o}rz}}},\ }\bibfield  {title} {\bibinfo {title}
  {{Estimating the two-particle $K$-matrix for multiple partial waves and decay
  channels from finite-volume energies}},\ }\href
  {https://doi.org/10.1016/j.nuclphysb.2017.09.014} {\bibfield  {journal}
  {\bibinfo  {journal} {Nucl. Phys.}\ }\textbf {\bibinfo {volume} {B924}},\
  \bibinfo {pages} {477} (\bibinfo {year} {2017})},\ \Eprint
  {https://arxiv.org/abs/1707.05817} {arXiv:1707.05817 [hep-lat]} \BibitemShut
  {NoStop}%
\bibitem [{\citenamefont {Guo}\ \emph {et~al.}(2013)\citenamefont {Guo},
  \citenamefont {Dudek}, \citenamefont {Edwards},\ and\ \citenamefont
  {Szczepaniak}}]{Guo:2012hv}%
  \BibitemOpen
  \bibfield  {author} {\bibinfo {author} {\bibfnamefont {P.}~\bibnamefont
  {Guo}}, \bibinfo {author} {\bibfnamefont {J.}~\bibnamefont {Dudek}}, \bibinfo
  {author} {\bibfnamefont {R.}~\bibnamefont {Edwards}},\ and\ \bibinfo {author}
  {\bibfnamefont {A.~P.}\ \bibnamefont {Szczepaniak}},\ }\bibfield  {title}
  {\bibinfo {title} {{Coupled-channel scattering on a torus}},\ }\href
  {https://doi.org/10.1103/PhysRevD.88.014501} {\bibfield  {journal} {\bibinfo
  {journal} {Phys. Rev. D}\ }\textbf {\bibinfo {volume} {88}},\ \bibinfo
  {pages} {014501} (\bibinfo {year} {2013})},\ \Eprint
  {https://arxiv.org/abs/1211.0929} {arXiv:1211.0929 [hep-lat]} \BibitemShut
  {NoStop}%
\bibitem [{\citenamefont {Blatt}\ and\ \citenamefont
  {Biedenharn}(1952)}]{Blatt:zz1952a}%
  \BibitemOpen
  \bibfield  {author} {\bibinfo {author} {\bibfnamefont {J.~M.}\ \bibnamefont
  {Blatt}}\ and\ \bibinfo {author} {\bibfnamefont {L.~C.}\ \bibnamefont
  {Biedenharn}},\ }\bibfield  {title} {\bibinfo {title} {{Neutron-Proton
  Scattering with Spin-Orbit Coupling. 1. General Expressions}},\ }\href
  {https://doi.org/10.1103/PhysRev.86.399} {\bibfield  {journal} {\bibinfo
  {journal} {Phys. Rev.}\ }\textbf {\bibinfo {volume} {86}},\ \bibinfo {pages}
  {399} (\bibinfo {year} {1952})}\BibitemShut {NoStop}%
\bibitem [{\citenamefont {{G. K\"all\'en}}(1964)}]{KallenBook}%
  \BibitemOpen
  \bibfield  {author} {\bibinfo {author} {\bibnamefont {{G. K\"all\'en}}},\
  }\href@noop {} {\emph {\bibinfo {title} {Elementary Particle Physics}}}\
  (\bibinfo  {publisher} {Addison-Wesley},\ \bibinfo {year} {1964})\BibitemShut
  {NoStop}%
\bibitem [{\citenamefont {Kuberski}(2023)}]{Kuberski:2023zky}%
  \BibitemOpen
  \bibfield  {author} {\bibinfo {author} {\bibfnamefont {S.}~\bibnamefont
  {Kuberski}},\ }\bibfield  {title} {\bibinfo {title} {{Low-mode deflation for
  twisted-mass and RHMC reweighting in lattice QCD}},\ }\href@noop {} {\
  (\bibinfo {year} {2023})},\ \Eprint {https://arxiv.org/abs/2306.02385}
  {arXiv:2306.02385 [hep-lat]} \BibitemShut {NoStop}%
\bibitem [{\citenamefont {Stanzione}\ \emph {et~al.}(2020)\citenamefont
  {Stanzione}, \citenamefont {West}, \citenamefont {Evans}, \citenamefont
  {Minyard}, \citenamefont {Ghattas},\ and\ \citenamefont {Panda}}]{frontera}%
  \BibitemOpen
  \bibfield  {author} {\bibinfo {author} {\bibfnamefont {D.}~\bibnamefont
  {Stanzione}}, \bibinfo {author} {\bibfnamefont {J.}~\bibnamefont {West}},
  \bibinfo {author} {\bibfnamefont {R.}~\bibnamefont {Evans}}, \bibinfo
  {author} {\bibfnamefont {T.}~\bibnamefont {Minyard}}, \bibinfo {author}
  {\bibfnamefont {O.}~\bibnamefont {Ghattas}},\ and\ \bibinfo {author}
  {\bibfnamefont {D.}~\bibnamefont {Panda}},\ }\bibfield  {title} {\bibinfo
  {title} {Frontera: The evolution of leadership computing at the national
  science foundation},\ }in\ \href@noop {} {\emph {\bibinfo {booktitle}
  {Proceedings of Practice and Experience in Advanced Research Computing (PEARC
  '20)}}}\ (\bibinfo {year} {2020})\BibitemShut {NoStop}%
\bibitem [{\citenamefont {Harris}\ \emph {et~al.}(2020)\citenamefont {Harris},
  \citenamefont {Millman}, \citenamefont {Van Der~Walt}, \citenamefont
  {Gommers}, \citenamefont {Virtanen}, \citenamefont {Cournapeau},
  \citenamefont {Wieser}, \citenamefont {Taylor}, \citenamefont {Berg},
  \citenamefont {Smith} \emph {et~al.}}]{harris2020array}%
  \BibitemOpen
  \bibfield  {author} {\bibinfo {author} {\bibfnamefont {C.~R.}\ \bibnamefont
  {Harris}}, \bibinfo {author} {\bibfnamefont {K.~J.}\ \bibnamefont {Millman}},
  \bibinfo {author} {\bibfnamefont {S.~J.}\ \bibnamefont {Van Der~Walt}},
  \bibinfo {author} {\bibfnamefont {R.}~\bibnamefont {Gommers}}, \bibinfo
  {author} {\bibfnamefont {P.}~\bibnamefont {Virtanen}}, \bibinfo {author}
  {\bibfnamefont {D.}~\bibnamefont {Cournapeau}}, \bibinfo {author}
  {\bibfnamefont {E.}~\bibnamefont {Wieser}}, \bibinfo {author} {\bibfnamefont
  {J.}~\bibnamefont {Taylor}}, \bibinfo {author} {\bibfnamefont
  {S.}~\bibnamefont {Berg}}, \bibinfo {author} {\bibfnamefont {N.~J.}\
  \bibnamefont {Smith}}, \emph {et~al.},\ }\bibfield  {title} {\bibinfo {title}
  {Array programming with numpy},\ }\href@noop {} {\bibfield  {journal}
  {\bibinfo  {journal} {Nature}\ }\textbf {\bibinfo {volume} {585}},\ \bibinfo
  {pages} {357} (\bibinfo {year} {2020})}\BibitemShut {NoStop}%
\bibitem [{\citenamefont {Hunter}(2007)}]{Hunter:2007}%
  \BibitemOpen
  \bibfield  {author} {\bibinfo {author} {\bibfnamefont {J.~D.}\ \bibnamefont
  {Hunter}},\ }\bibfield  {title} {\bibinfo {title} {Matplotlib: A 2d graphics
  environment},\ }\href {https://doi.org/10.1109/MCSE.2007.55} {\bibfield
  {journal} {\bibinfo  {journal} {Computing in Science \& Engineering}\
  }\textbf {\bibinfo {volume} {9}},\ \bibinfo {pages} {90} (\bibinfo {year}
  {2007})}\BibitemShut {NoStop}%
\bibitem [{\citenamefont {Edwards}\ and\ \citenamefont
  {Joo}(2005)}]{Edwards:2004sx}%
  \BibitemOpen
  \bibfield  {author} {\bibinfo {author} {\bibfnamefont {R.~G.}\ \bibnamefont
  {Edwards}}\ and\ \bibinfo {author} {\bibfnamefont {B.}~\bibnamefont {Joo}}
  (\bibinfo {collaboration} {SciDAC}),\ }\bibfield  {title} {\bibinfo {title}
  {{The Chroma software system for lattice QCD}},\ }\href@noop {} {\bibfield
  {journal} {\bibinfo  {journal} {Nucl. Phys. Proc. Suppl.}\ }\textbf {\bibinfo
  {volume} {140}},\ \bibinfo {pages} {832} (\bibinfo {year}
  {2005})}\BibitemShut {NoStop}%
\end{thebibliography}%
\end{document}